\documentclass[a4paper,11pt]{article}

\usepackage{amssymb,amsmath,amsfonts,epsfig}
\usepackage{tikz}
\usetikzlibrary{shapes, arrows, shadows}

\def\be{\begin{equation}}
\def\ee{\end{equation}}
\def\bea{\begin{eqnarray}}
\def\eea{\end{eqnarray}}

\begin{document}
\begin{titlepage}
\date{today}       \hfill

\begin{center}

\vskip .5in
{\Large \bf   Fusion of conformal interfaces and bulk induced boundary RG flows}\\
\vspace{5mm}

\today 
 
\vskip .250in

\vskip .5in
{\large Anatoly Konechny}

\vskip 0.5cm
{\it Department of Mathematics,  Heriot-Watt University\\
Edinburgh EH14 4AS, United Kingdom\\[10pt]
and \\[10pt]
Maxwell Institute for Mathematical Sciences\\
Edinburgh, United Kingdom\\[10pt]
}
E-mail: A.Konechny@hw.ac.uk
\end{center}

\vskip .5in
\begin{abstract} \large
We consider the basic radius changing  conformal interface for a free compact boson. 
After investigating different theoretical aspects of this object we 
focus on the fusion of this interface
%
with conformal boundary conditions. At fractions of the self-dual 
radius there exist exceptional D-branes. It was argued in \cite{Gaberdiel_etal} that changing the radius in the bulk 
induces a boundary RG flow. Following \cite{BR} we conjecture that fusing the basic radius changing interface (that changes the radius 
from a fraction of the self-dual radius) with the exceptional boundary conditions 
gives the boundary condition which is the end point of the RG flow considered 
in \cite{Gaberdiel_etal}.  By studying the fusion singularities we recover RG logarithms and see, in particular instances,  how they get 
resummed into power singularities. We discuss what quantities need to be calculated to gain full non-perturbative 
control over the fusion.

\end{abstract}

\end{titlepage}

\renewcommand{\thepage}{\arabic{page}}
\setcounter{page}{1}
\large 

\section{Introduction}
\renewcommand{\theequation}{\arabic{section}.\arabic{equation}}

Conformal interfaces are one-dimensional objects that separate two  two-dimensional conformal field theories: ${\rm CFT}_1$ 
and ${\rm CFT}_2$. Conformal interfaces specify operators 
\be\label{general_interface}
{\cal O}_{21}: {\cal H}_1\to {\cal H}_2^{*}
\ee
 acting between the state spaces of the theories\footnote{We put the dual space 
${\cal H}^{*}_2$ as the target space because the images of conformal interfaces often have infinite norm but have finite overlaps with vectors 
from ${\cal H}_2$.}. The interface being conformal implies the relations 
\be
(L_{n}^{(2)}-\bar L_{-n}^{(2)}){\cal O}_{21} = {\cal O}_{21}(L_{n}^{(1)}-\bar L_{-n}^{(1)})
\ee
where $L^{(i)}_{n}$ and $\bar L_{n}^{(i)}$ are the left and right Virasoro algebra modes in the corresponding theories. 
If it happens that ${\cal O}_{21}$ satisfies stronger relations 
\be
L_{n}^{(2)}{\cal O}_{21} = {\cal O}_{21}L_{n}^{(1)}\, , \qquad
\bar L_{n}^{(2)}{\cal O}_{21} = {\cal O}_{21}\bar L_{n}^{(1)}
\ee
the corresponding interface is called topological. Such interface can be moved freely in space without changing any correlation functions. 

Folding along the interface line we obtain an alternative picture \cite{AO} in which the conformal interface is described as 
a conformal boundary condition in the tensor product ${\rm CFT}_{1}\otimes {\rm CFT}_2$. In the case when 
${\rm CFT}_1$ is the same as ${\rm CFT}_2$ the interface is called a defect\footnote{Often the interfaces between different CFT's are also called  defects. 
In this paper we will use both terms judiciously.}. A conformal boundary condition in  ${\rm CFT}_1$ can be considered as an interface between 
${\rm CFT}_1$ and a trivial CFT (whose state space contains only the vacuum).  

Given a conformal interface (\ref{general_interface}) we can consider its fusion with a conformal boundary condition $|B\rangle\!\rangle_{1}\in {\cal H}^{*}_1$ in 
${\rm CFT}_1$. 
We put the interface on a semi-infinite cylinder placing it distance $\epsilon$ away from the end which is capped by the boundary state
 $|B\rangle\!\rangle$ (see figure 1 below). We then send $\epsilon$ to zero 
subtracting a divergence:
\be\label{fuse_naive}
|{\cal O}_{21} \circ B\rangle\!\rangle_{2} = \lim_{\epsilon \to 0} e^{d/\epsilon} {\cal O}_{21} e^{-\epsilon H_2} |B\rangle\!\rangle_{1}
\ee

\begin{center}
\begin{tikzpicture}[>=latex]
\draw[thick] (-2,0)--(6.4,0);
\draw[thick] (-2,2.2)--(6.4,2.2);

\draw[thick, dashed] (-2,0) arc (270:90:0.6 and 1.1);
\draw[thick, dashed] (-2,0) arc (-90:90:0.6 and 1.1 );
\shadedraw[fill=black!10!white,draw=black, very thick] (6.4,1.1) circle [x radius=0.6, y radius = 1.1];

\draw[very thick] (4,0) arc (270:90:0.6 and 1.1);
\draw[very thick, dashed] (4.0,0) arc (-90:90:0.6 and 1.1);

\draw (7.5,1.1) node {$|B\rangle\!\rangle$};
\draw (4.2,-0.3) node {${\cal O}_{21}$};
\draw (1.3,1.1  ) node {${\rm CFT}_2$};
\draw (5.2,1.1) node {${\rm CFT}_1$};
\draw (4,2.25) -- (4, 2.6);
\draw (6.4,2.25)--(6.4,2.6);
\draw[<->] (4, 2.45) -- (6.4,2.45);
\draw (5.2,2.64) node {$\epsilon$};
\end{tikzpicture}
\vspace{5mm}

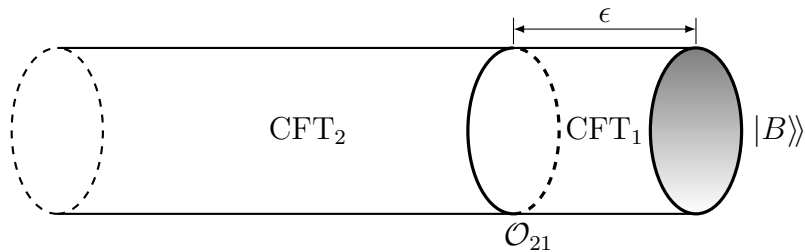
{\normalsize {\bf Figure 1}: Fusion of a conformal interface with a boundary state.}
\end{center}
The result of the fusion is a conformal boundary state $|{\cal O}_{21} \circ B\rangle\!\rangle_{2}$ in ${\rm CFT}_2$.
In (\ref{fuse_naive}) we assume that the $\epsilon\to$ divergence takes the form of an overall divergent factor 
$e^{-d/\epsilon}$ where $d$ is some constant whose role is similar to Casimir energy between two boundary conditions separated by 
distance $\epsilon$.
We will discuss  the divergences associated with fusion in more detail 
later. 

There is a number of interesting connections between interfaces and renormalisation group (RG) flows explored in the literature.  
In \cite{GW} it was shown that topological defects act on boundary RG flows. In \cite{BG} special topological defects were constructed whose fusion
with  a boundary condition that specifies a UV fixed point of a boundary RG flow gives the corresponding IR boundary condition. 
For the Kondo model flow the corresponding defect operator is a renormalised loop operator.  
RG flows are triggered by perturbations. 
It was proposed in \cite{FQ}, \cite{BR}  to look at interfaces obtained by perturbing the ultraviolet  ${\rm CFT}_{\rm UV}$  on a half plane. 
This may trigger  an RG 
flow on the interface itself. Following  the resulting bulk plus boundary RG flow we obtain a particular conformal interface between 
${\rm CFT}_{\rm UV}$  and ${\rm CFT}_{\rm IR}$. 
In \cite{Gaiotto} an algebraic construction  of such an interface was put forward 
for the $\psi_{13}$ - flows between neighbouring minimal models \cite{Zam}. Such RG (or perturbation) interfaces must contain information about the bulk RG flows. 
Moreover, it was argued in \cite{BR} that fusing a bulk RG interface with a boundary condition $|B\rangle\!\rangle$ in ${\rm CFT}_{\rm UV}$ gives the end point of a {\it boundary}  RG flow triggered by the same bulk perturbation on $|B\rangle\!\rangle$.  In \cite{BR} this proposal was 
tested for $N=2$ supersymmetric flows between minimal models and in \cite{BR2}, \cite{Bachas_BR} further examples of supersymmetric flows were studied.
 Analogues of the RG interfaces of  \cite{FQ},  \cite{BR} 
for pure boundary RG flows were proposed 
 in \cite{Konechny} where it was argued that they are represented by particular boundary condition changing operators.  

 The above relations thus concern pure boundary, pure bulk and coupled bulk plus boundary RG flows. 
In the present paper we are interested in a version of the proposal 
of \cite{BR} in which the bulk perturbation is exactly marginal but it does trigger a non-trivial boundary RG flow. 
We next discuss the general picture of such flows in more detail. 


Suppose the couplings $\lambda^{I}$ are all exactly marginal bulk couplings that couple to operators $\phi_{I}$. 
We thus have a family of bulk CFT's: ${\rm CFT}_{\lambda^{I}}$.  
Consider a conformal boundary condition with a boundary state $|B\rangle\!\rangle_{0}$ defined in the ${\rm CFT}_{0}$. 
If we perturb the bulk theory ${\rm CFT}_{0}$ by a linear combination $\lambda^{I}\phi_{I}$ we may get additional divergences 
arising from collisions of operators $\phi_{I}$ with the boundary. At the leading order the divergence comes from a bulk-to-boundary 
operator product expansion (OPE) 
\be
\phi_{I}(x,\tau) \sim \frac{1}{(2x)^{\Delta_{I} - \Delta_{i}}} B_{I}^{i}\psi_{i}(\tau) \, . 
\ee
Here $x$ is the coordinate transverse to the boundary, $\tau$ is the coordinate along the boundary, 
$\psi^{i}$ are boundary scaling fields in the theory specified by $|B\rangle\!\rangle_{0}$. The coefficients 
$B_{I}^{i}$ are the bulk-to-boundary OPE coefficients.
When the difference of bulk and boundary scaling dimensions $\Delta_{I} - \Delta_{i}$ is greater than 1 we 
have a perturbation theory divergence near $x=0$. When  $\Delta_{I} - \Delta_{i}=1$ the divergence is 
logarithmic and results in a universal term in the beta function for the boundary couplings $\mu^{i}$ that 
couple to $\psi^{i}$:
\be\label{leading_beta}
\beta^{i} = \frac{1}{2}B^{i}_{I}\lambda^{I} + \dots 
\ee
We imagine  constructing a non-conformal boundary condition $|B\rangle\!\rangle_{\lambda^{I}}^{\rm UV}$ in the 
deformed bulk CFT: ${\rm CFT}_{\lambda^{I}}$ by subtracting all boundary divergences by the corresponding boundary counter terms.  
These counterterms depend on $\lambda^{I}$ and will result in a beta function for the boundary couplings of the form 
 \be\label{beta_general}
 \beta^{i} = \beta^{i}_{(0)}(\lambda) + \mu^{j}\beta_{(1)j}^{i}(\mu, \lambda)\, .
 \ee
 Here the first term 
 \be \label{beta_general2}
 \beta^{i}_{(0)}(\lambda) = \frac{1}{2}B^{i}_{I}\lambda^{I} + B^{i}_{IJ} \lambda^{I}\lambda^{J} + \dots 
 \ee
contains $\mu^{i}$-independent terms which correspond to divergences  arising from collisions of the bulk operators with the boundary 
specified by $|B\rangle\!\rangle_{0}$. For example the term proportional to $\lambda^{I}\lambda^{J}$ comes from simultaneous 
collisions of $\phi_{I}$ and $\phi_{J}$ at the boundary. While the leading term specified by the bulk-to-boundary OPE is universal 
the higher order terms in $\beta^{i}_{(0)}$ depend on the subtraction scheme. The boundary condition $|B\rangle\!\rangle_{\lambda^{I}}^{\rm UV}$ 
flows under the RG specified by (\ref{beta_general}). We flow out of the $\mu^{i}=0$ point along the tangent vector specified by 
$\beta^{i}_{(0)}(\lambda) $. The end point of the flow is some conformal boundary condition $|B\rangle\!\rangle_{\lambda^{I}}^{\rm IR}$ 
in ${\rm CFT}_{\lambda^{I}}$ which may depend on  $\beta^{i}_{(0)}(\lambda) $ and thus {\it on the subtraction scheme} that specifies this part of the 
beta function.  
While the g-theorem \cite{AL1}, \cite{FK} certainly applies along this boundary RG flow it is not of much use for us
 because we do not know from what value of the boundary entropy does the flow start\footnote{
The author is much indebted to Daniel Friedan for illuminating discussions about this issue.}. 
This is the value of the boundary entropy $s_{\rm UV}$ for  $|B\rangle\!\rangle_{\lambda^{I}}^{\rm UV}$ which in general we 
have no  control over and which is different from the boundary 
entropy $s_{0}$ of $|B\rangle\!\rangle_{0}$. 


If we treat the bulk deformation {\it infinitesimally }then the flow is given by all terms in $\beta^{i}$ linear in $\lambda^{I}$. This 
part of the beta function is free from  ambiguities. 
The value of the boundary entropy $s_{\rm UV}$  is only infinitesimally different from $s_{0}$ 
and since during the RG flow it changes a finite amount we can conclude that (for infinitesimal 
bulk deformations) the boundary entropy of $|B\rangle\!\rangle_{\lambda^{I}}^{\rm IR}$ must be smaller than $s_{0}$ 
which is certainly a useful constraint.

In this paper we are specifically interested in the bulk induced boundary flow studied in \cite{Gaberdiel_etal}. The authors consider 
a free compact boson theory. The conformal boundary conditions in this theory were classified in \cite{Friedan} 
(see also \cite{GRW}, \cite{GR}, \cite{Janik}). For a generic radius of compactness the only irreducible conformal boundary 
conditions with finite boundary entropy are the Dirichlet and Neumann ones. If the radius is a rational fraction of the self-dual radius 
there are additional conformal boundary conditions labelled by points on the $SU(2)$ group manifold modded out by certain discrete 
 subgroups \cite{Callan_etal}, \cite{RS}, \cite{GRW}, \cite{GR}. In particular for the self-dual radius the general irreducible boundary condition (b.c.) is specified by an SU(2) group element. 
 The Dirichlet and Neumann boundary conditions are contained in the SU(2) group manifold as two non-intersecting circles. 
 These two circles are preserved by the bulk radius deformation. 
 If we take a boundary condition away from these two circles and deform the bulk radius we get a boundary RG flow.
 In \cite{Gaberdiel_etal} these RG flows were studied  infinitesimally 
 in the bulk deformation. The g-theorem then predicts that (unless we initially have a Neumann b.c.) increasing the radius will trigger a flow that ends up at a single Dirichlet boundary condition 
 while  decreasing the radius we should end up at a single Neumann b.c. (unless we started from a Dirichlet one).   
 In \cite{Gaberdiel_etal} the beta function was calculated at the linear order in the bulk coupling and to all orders in the boundary couplings.  
 The resulting flow on the group manifold confirmed the predictions of the g-theorem. Note that there is always an instability (or discontinuity)
 at one of the special circles, e.g. when we increase the radius the circle of Neumann boundary conditions remains intact but 
 any point near it will flow to a Dirichlet boundary condition. 
 
 In this paper we study the fusion of the basic radius-changing interface in the free boson theory  with the exceptional boundary states 
 $|g\rangle\!\rangle$, $g\in {\rm SU(2)}$ at the self-dual radius. This interface, which will be described in  detail in section 2, can 
 be obtained as a perturbation interface. We consider  the free boson theory at the self-dual radius  put on a semi-infinite cylinder 
 with the boundary state put at the $\tau=0$ end. We then perturb part of the infinite cylinder $\tau<-\epsilon$ by the radius changing 
 operator $\partial \phi \bar \partial \phi$
 as depicted below

\begin{center}
\begin{tikzpicture}[>=latex]

\fill[red!10!white] (-2,0) -- (4,0) arc    (-90:90:0.6 and 1.1 ) -- (-2,2.2) arc (90:270: 0.6 and 1.1)--cycle;
\draw[thick] (-2,0)--(6.4,0);
\draw[thick] (-2,2.2)--(6.4,2.2);


\draw[thick, dashed] (-2,2.2) arc (90: 270: 0.6 and 1.1);
\draw[thick, dashed] (-2,0) arc (-90:90:0.6 and 1.1 );
\shadedraw[fill=black!10!white,draw=black, very thick] (6.4,1.1) circle [x radius=0.6, y radius = 1.1];

\draw[very thick] (4,0) arc (270:90:0.6 and 1.1);
\draw[very thick, dashed] (4.0,0) arc (-90:90:0.6 and 1.1);

\draw (7.5,1.1) node {$|g\rangle\!\rangle$};
\draw (4.6,-0.3) node {${\cal I}^{(R\leftarrow R_{\rm s.d.})}$};
\draw (1.3,1.1  ) node {$e^{\lambda\! \int\! \partial \phi \bar \partial \phi}$};
\draw (1.2, 0.6) node {\tiny perturbed theory};
\draw (5.2,1.1) node {$R_{\rm s.d.}$};
\draw (4,2.25) -- (4, 2.6);
\draw (6.4,2.25)--(6.4,2.6);
\draw[<->] (4, 2.45) -- (6.4,2.45);
\draw (5.2,2.64) node {$\epsilon$};

\end{tikzpicture}
\vspace{5mm}

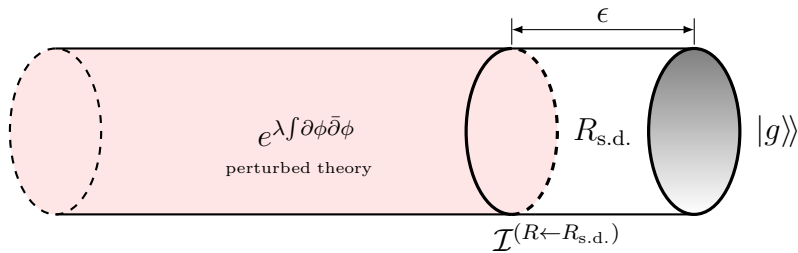
{\normalsize {\bf Figure 2}: Fusion of the radius changing interface with the exceptional  boundary state.}
\end{center}

The interface operator ${\cal I}^{(R\leftarrow R_{\rm s.d.})}$ we choose corresponds to  a particular renormalisation prescription for the perturbation theory singularities 
that resolves collisions of the radius changing operators in the bulk (away from $\tau=-\epsilon$) 
 and at the position of the interface: $\tau=-\epsilon$. 
The coupling constant $\lambda$ is related to the value of the radius of compactness to the left of the interface
(see formula (\ref{lambda1}) below). The distance $\epsilon$ serves as a regulator  for divergences near the boundary at $\tau=0$. 
Sending $\epsilon$ to zero we will have to deal with new singularities. We see that the fusion singularities correspond to 
perturbation theory singularities treated in a particular way -- we first deal with all collisions of the bulk operators away 
from the boundary (that is summarised in the interface operator) and then we treat the collisions with the boundary by 
sending  $\epsilon$ to zero and subtracting the fusion divergences. In section \ref{sec_pert} we will make this connection 
more direct  by showing how 
the RG logarithms appear as  fusion divergences. We conjecture then that the results of fusion of ${\cal I}^{(R\leftarrow R_{\rm s.d.})}$ 
with the boundary state $|g\rangle\!\rangle $ is the end point of the boundary RG flow triggered by the bulk marginal 
perturbation with a particular   $\beta_{(0)}^{i}(\lambda)$. The scheme for dealing with multiple collisions near the boundary is 
fixed via  the interface operator ${\cal I}^{(R\leftarrow R_{\rm s.d.})}$ and the regulator $\epsilon$. The use 
of the interface operator in principle, if not in practice,  allows us  to  study the flow for a {\it finite} value of the bulk coupling.
For a generic value of $R$ the available fixed points are superpositions of Dirichlet and Neumann branes. 
While we are not aware of an argument that would exclude a non-trivial superposition it seems to us most plausible 
that the end point of the RG flow for a finite deformation is the same as for the infinitesimal one considered in 
\cite{Gaberdiel_etal}. Our calculations presented in this paper support this conjecture.

To have control over the fusion process we need to have some idea of what type of singularities to expect. 
In general  we expect the outcome of a fusion of an interface ${\cal O}_{21}$  with a boundary state  $|B\rangle\!\rangle_{1}$ 
to be a boundary state which can be decomposed into Ishibashi states $|I\rangle\!\rangle_{2}$ so that
\be\label{fusion_structure1}
{\cal O}_{21} e^{-\epsilon H_2} |B\rangle\!\rangle_{1} = \sum_{I} {\cal A}_{I}(\epsilon) |I\rangle\!\rangle_{2} + \dots 
\ee
where ${\cal A}_{I}(\epsilon) $ are some functions of $\epsilon$ which are singular at $\epsilon =0$ and the ellipsis stands 
for the terms vanishing in the limit $\epsilon \to 0$.  In particular among the Ishibashi states $|I\rangle\!\rangle_{2}$ 
we have the one corresponding to the vacuum state $|0\rangle\!\rangle_{2}$. We expect the corresponding amplitude 
${\cal A}_{0}$ to have an essential singularity 
\be
{\cal A}_{0}(\epsilon) \sim g_{0} e^{-\frac{{\cal E}_{0}}{\epsilon}} 
\ee
where $g_{0}$ and ${\cal E}_{0}$ are constants. The reason for this is explained in \cite{BBDO} (see section 2.2 and Appendix A in 
particular). We can cut off the semi-infinite cylinder at $\tau=-L$ where we can put some local boundary condition. 
Sending $\epsilon$ to zero can then be viewed in the "open string" channel in which we quantise along the periodic  direction $\sigma$.
In that quantisation we have two local boundary conditions at $\tau=0,-L$ and the interface put at $\tau=-\epsilon$. 
The energy levels are discrete as long as $L$ is finite. 
The leading singularity at $\epsilon \to 0$ will come from the ground state energy ${\cal E}_{0}$ in this channel which we can call 
the fusion Casimir energy. The dependence on the choice of the second boundary condition at $\tau=-L$ should disappear in the 
$L\to \infty$ limit (the part of the Casimir energy independent of the interface ${\cal O}_{21}$ goes as $1/L$ and vanishes in the limit).

The same general reasoning however does not work for the amplitudes ${\cal A}_{I}(\epsilon)$ in other sectors.  
In particular for  the fusion ${\cal I}^{(R \leftarrow R_{\rm s.d.} )}\circ |g \rangle\!\rangle$ 
the fusion looks like
    \be\label{fusion_structure2}
  {\cal I}^{(R \leftarrow R_{\rm s.d.})} q^{2L_{0}} |g \rangle\!\rangle \sim C_0 e^{-{\cal E}_{0}/\epsilon} |0\rangle\!\rangle_{R}
  +  \sum_{N\ne 0} A_{N}(\epsilon)e^{-{\cal E}_{N}^{\rm D}/\epsilon} |N,0\rangle\!\rangle_{R} 
  +  \sum_{M\ne 0} B_{M}(\epsilon) e^{-{\cal E}_{M}^{\rm N}/\epsilon} |0,M\rangle\!\rangle_{R} 
    \ee
  where $|0\rangle\!\rangle_R$, $|N,0\rangle\!\rangle_{R} $, $|0,M\rangle\!\rangle_{R} $ stand for the vacuum, Dirichlet-type and 
  Neumann-type Ishibashi states respectively. We have singled out the essential singularities with the functions 
  $A_{N}(\epsilon)$, $B_{N}(\epsilon)$ containing possible milder singularities or zeroes. 
  
  Note that for $N\ne$ or $M\ne 0$, unlike in the vacuum case, we cannot ensure picking the correct momentum or  
  winding sector by putting a {\it local} boundary condition at $\tau =-L$ end\footnote{We assume the boundary 
  condition at $\tau=-L$ must be  $L$-independent, otherwise we could impose having  momentum $p$ at infinity by 
  requiring the fall off 
  $\phi(\tau) \sim -p \ln(-\tau)$}. As there are no known examples in which the fusion amplitudes are calculated exactly and there is 
  a non-trivial RG flow, it is not clear to us what kind of behaviour to expect for $A_{N}(\epsilon)$, $B_{N}(\epsilon)$. 
  Part of the motivation for the present project is to investigate the general structure of  (\ref{fusion_structure1}).
  Regardless of these unknown singularities we expect the vacuum Ishibashi state $|0\rangle\!\rangle_{2}$ 
  to be present in the fused boundary condition and therefore the subtraction given in (\ref{fuse_naive}) with $d={\cal E}_{0}$ still looks 
  reasonable. 
    
  The rest of this paper is organised as follows. 
  In section \ref{section_interface} we discuss the basic radius changing  interface ${\cal I}^{(R_1 \leftarrow R_2)}$ from different perspectives  
   - as a conformal interface, as an operator 
implementing a Bogolyubov transformation, as a perturbation interface  and as an operator defining transport of 
operators under marginal deformation. In section \ref{section_fusion1} we discuss fusion of ${\cal I}^{(R_1 \leftarrow R_2)}$ with 
conformal boundary conditions. After working out the simple cases of  Neumann and Dirichlet branes as a warm up, 
we turn to fusion with exceptional boundary states. We work out what basic set of amplitudes  needs to be computed to determine the fusion 
and find their representations in terms of traces of certain operators in chiral Fock spaces. 
In section \ref{sec_pert} we present various perturbative calculations for these basic amplitudes. In section \ref{C1} we calculate 
non-perturbatively the vacuum fusion  amplitude ${\cal A}_{0}(\epsilon)$ for  $R=\infty$. We conclude with some brief comments in 
section \ref{conclusions}. Some technical details are put into two appendices.
  

\section{The radius changing interface}\label{section_interface}
\setcounter{equation}{0}

We consider a free compact boson $\phi(x,\tau)$ in  two-dimensional space-time 
with action 
\be\label{free_action}
S = \frac{1}{8\pi} \int\!\! d\sigma\! \! \int\!\! dt ( (\partial_{t} \phi)^2 - (\partial_{x} \phi)^2 ) \, .
\ee
We have identification $\phi \sim \phi + 2\pi R $ where $R$ is the radius of compactness. 
We consider this theory on a  circle of circumference $2\pi$ so that $\sigma \sim \sigma + 2\pi$ 
is the periodic coordinate and $t$ is the (real) time variable. The mode expansion 
is
\bea\label{modes}
&& \phi(\sigma, t) = \phi_0 +p_{L}(t - \sigma) +p_{R}(t+\sigma)    \nonumber \\
&& + \sum_{n=1}^{\infty} \frac{i}{\sqrt{n} }\Bigl[ e^{-int}(a_n e^{in\sigma} 
+ \bar a_n e^{-in\sigma}) - e^{int} (a_n^{\dagger} e^{-in\sigma} + \bar a_{n}^{\dagger}e^{in\sigma})\Bigr]
\eea 
where    the oscillator modes satisfy 
\be
[a_n, a^{\dagger}_{m}]=\delta_{n,m} \, , \qquad [\bar a_n, \bar a^{\dagger}_{m}]=\delta_{n,m} \, .
\ee
Going to Euclidean time $\tau=it$ we obtain a theory on an infinite cylinder with $\tau$ being the coordinate along its axis. 

The zero modes $p_{L}$, $p_{R}$ are quantised as
\be
p_{L} = \frac{N}{R} + \frac{MR}{2}\, , \qquad p_{R}= \frac{N}{R} -\frac{MR}{2} 
\ee
where $N\in {\mathbb Z}$ and $M\in {\mathbb Z}$ are momentum and winding quantum numbers.
We denote the corresponding normalised $U(1)$ primary states as $|N,M \rangle_{R}$ and the Fock spaces 
built on them as ${\cal F}_{N,M;R}$. The complete state space is 
\be
{\cal H}_{R} = \bigoplus_{N,M} {\cal F}_{N,M;R} \, .
\ee

Conformal interfaces between two free boson CFT's with radii $R_1$ and $R_2$ that preserve the $U(1)$ symmetry 
were studied in \cite{BB}. They are represented either as operators 
\be \label{defect_op}
{\cal I}(1\leftarrow 2): {\cal H}_{R_{2}}\to {\cal H}_{R_{1}} \, ,
\ee
or in the folded picture (\cite{AO}, \cite{BBDO}) they can be described  as D1-branes on a square two-torus with radii 
$R_1$, $R_2$.  Such D1-branes are parameterised by two winding numbers and two Wilson line parameters. 
The basic radius changing interface is given by the D1 brane that winds around each basic cycle once and has trivial 
Wilson lines. This interface is called the deformed identity interface in \cite{BB}. 
The corresponding operator (\ref{defect_op}) 
is 
\be \label{I}
{\cal I}^{(R_1\leftarrow R_2)} =g_{\vartheta} \sum_{N,M \in {\mathbb Z}} |N,M\rangle_{R_1} \langle N,M|_{R_2} \prod_{n=1}^{\infty} \exp\Bigl[ C(a_n\bar a_{n} 
- b_{n}^{\dagger}\bar b_{n}^{\dagger}) +S(b_{n}^{\dagger}a_n + \bar b_{n}^{\dagger}\bar a_n)\Bigr]
\ee
Here $a_{n}, \bar a_n$ are the annihilation operators of ${\cal H}_{R_2}$ and $b_n^{\dagger}, \bar b_{n}^{\dagger}$ are 
the creation modes of ${\cal H}_{R_1}$ which in  (\ref{I}) are understood to be acting on $ |N,M\rangle_{R_1}$ from the left. 
The coefficients $C,S,g_{\vartheta}$ are 
\be
C = \frac{(R_1)^2 - (R_2)^2}{(R_1)^2 + (R_2)^2} \,  , \qquad S= \frac{2R_1 R_2}{(R_1)^2 + (R_2)^2} \, , \qquad 
g_{\vartheta} = \frac{1}{\sqrt{S}} \, . 
\ee
They can be expressed in terms of $\vartheta$ - the angle which the corresponding diagonally stretched D1-brane forms with a side of the two-torus 
\be
\vartheta = \arctan\left( \frac{R_2}{R_1}\right) \, , \qquad C= \cos(2\vartheta)\, , \qquad S=\sin(2\vartheta) \, .
\ee 
The overall coefficient $g_{\vartheta}$ is the Affleck and Ludwig's $g$-factor \cite{AL1} of ${\cal I}^{(R_1\leftarrow R_2)}$.


\subsection{${\cal I}^{(R_1\leftarrow R_2)}$ and  Bogolyubov transformation}
A change of radius for a compact boson can be implemented by a Bogolyubov transformation. For 
the oscillator modes we have \cite{KZ}
\bea \label{Bog1}
a_n'& = &\cosh(\chi) a_n - \sinh(\chi) \bar a_n^{\dagger} \, , \nonumber \\
\bar a_n'& = &\cosh(\chi) \bar a_n - \sinh(\chi)  a_n^{\dagger}
\eea 
where $a_n, \bar a_n$ correspond to the radius $R_2$ and $a_n',\bar a_n'$ correspond to $R_1$, and
\be
\cosh(\chi) = \frac{(R_{1})^2 + (R_2)^2}{2R_1 R_2}\, , \qquad \sinh(\chi) = \frac{(R_1)^2 - (R_2)^2}{2R_1R_2} \, .
\ee
The zero modes $p_L,p_R$ are rotated as
\bea\label{Bog2}
\phi_{0}'&=& \phi_{0} \frac{R_1}{R_2} = e^{\chi} \phi_{0} \nonumber \\
p_L'&=&\cosh(\chi) p_L  - \sinh(\chi) p_R \, , \nonumber \\ 
p_R'&=&\cosh(\chi) p_R - \sinh(\chi) p_L
\eea
where again the primed quantities  correspond to $R_1$ and the unprimed ones to $R_2$. 
The last two relations mean that 
the winding and momentum integers: $N$, $M$  are invariant. 
The above identities between the modes stem from the gluing conditions on the fields $\phi$ and $\phi'$ set at 
$t=0$:
\be
\frac{\phi(\sigma, 0)}{R_{2}} = \frac{\phi'(\sigma, 0)}{R_{1}} \, , \qquad 
R_{2}\partial_{t}\phi(\sigma, 0)  = R_1 \partial_{t}\phi'(\sigma, 0) \, .
\ee
In the folded picture these gluing conditions describe a D-brane on a two-torus stretched diagonally. 
It is clear then that the interface ${\cal I}^{(R_{1}\leftarrow R_2)}$ should encode the Bogolyubov transformation (\ref{Bog1}), (\ref{Bog2}).
It can be  checked directly  that the operator ${\cal I}^{(R_1\leftarrow R_2)}$ given in (\ref{I}) satisfies the relations 
\bea
& b_{n}^{\dagger} {\cal I}^{(R_1\leftarrow R_2)} = {\cal I}^{(R_1\leftarrow R_2)} (a_n')^{\dagger} \, , \qquad 
&\bar b_{n}^{\dagger} {\cal I}^{(R_1\leftarrow R_2)}  = {\cal I}^{(R_1\leftarrow R_2)} (\bar a_n')^{\dagger} \, , \nonumber \\
& b_{n} {\cal I}^{(R_1\leftarrow R_2)} = {\cal I}^{(R_1\leftarrow R_2)} a_n' \, , \qquad 
&\bar  b_{n}  {\cal I}^{(R_1\leftarrow R_2)} = {\cal I}^{(R_1\leftarrow R_2)} \bar a_n' \, .
\eea
The analogous relations for the zero modes also hold as ${\cal I}^{(R_1\leftarrow R_2)}$ preserves the quantum 
numbers $N, M$. Thus ${\cal I}^{(R_1\leftarrow R_2)}$ realises the Bogolyubov transformation (\ref{Bog1}), (\ref{Bog2}) as 
an intertwiner of  the Heisenberg algebras. 

We next want to understand the boundary entropy $g_{\vartheta}$ that is present in  ${\cal I}^{(R_1\leftarrow R_2)}$  as 
an overall normalisation factor
from the point of view  of Bogolyubov transformations.
 We first remind the reader some basic facts. 
Let $a_{i}^{\dagger}$, $a_{j}$ be a collection (possibly infinite) of creation and annihilation  operators satisfying the canonical commutation relations
$$[a_i,a^{\dagger}_j]=\delta_{i,j}\, .$$
A homogeneous canonical transformation can be written as 
\bea\label{gen_can}
a'_{i} &=& \sum_{j} (\Phi_{ij}a_j + \Psi_{ij}a^{\dagger}_j) \, , \nonumber \\
(a'_i)^{\dagger}&=& \sum_{j} (\Phi_{ij}^{*}a_j + \Psi_{ij}^{*}a^{\dagger}_j) \, .
\eea
This transformation is called proper (or unitarily realisable) if there exists a unitary operator $U$ such that 
\be\label{UUU}
a_i'=U a_i U^{-1} \, , \qquad (a_{i}')^{\dagger} = U a_i^{\dagger} U^{-1} \, .
\ee
Proper canonical transformations are usually called Bogolyubov transformations. 
It is known that (\ref{gen_can}) is proper if and only if the operator $\Psi$ is Hilbert-Schmidt (see e.g. \cite{Berezin}). In this case 
the operator $\Phi \Phi^{*}$ has the Fredholm determinant and the unitary operator $U$ can be obtained from 
the matrix form generating functional 
\be
\tilde U(a, a^{*}) = \frac{\theta}{({\rm det}\Phi\Phi^{\dagger})^{1/4}}\exp\Bigl[ \frac{1}{2}(a,a^{*}) 
\left(\begin{array}{rr} 
\Psi^{*}\Phi^{-1}&  (\Phi^{-1})^{T} \\
\Phi^{-1}& - \Phi^{-1}\Psi
\end{array}
\right)
\left(\! \begin{array}{l}
a\\
a^{*}
\end{array} \!\! \right)
\Bigr]
\ee
where $\theta$ is an arbitrary phase (see \cite{Berezin} formula (4.26)). 
Here for brevity we denoted by $a$ the vector $(a_1, a_2, \dots )$ and by $a^{*}$ the vector $(a_1^{*}, a_{2}^{*}, \dots)$. 
Recall that if the symbol $\tilde U(a, a^{*})$ is expanded as 
\be
\tilde U(a, a^{*})= \sum_{i_1, i_2, \dots , j_1, j_2, \dots} U_{i_1, i_2, \dots, j_1, j_2, \dots} a_{i_1}^{*}a_{i_2}^{*}\dots a_{j_1}a_{j_2} \dots
\ee
the corresponding operator can be written in terms of the creation and annihilation operators as 
\be
U =  \sum_{i_1, i_2, \dots , j_1, j_2, \dots} U_{i_1, i_2, \dots, j_1, j_2, \dots} a_{i_1}^{\dagger}a_{i_2}^{\dagger }\dots |0\rangle \langle 0| a_{j_1}a_{j_2} \dots
\ee

Bogolyubov transformation (\ref{Bog1}) for  fixed $n$ operates on  $a_n, \bar a_n, a_n^{\dagger}, \bar a_{n}^{\dagger}$,  
and is implemented by a unitary operator $U_n$ with a symbol 
\be\label{Un}
\tilde U_{n} = \frac{1}{\cosh(\chi)}\exp\Bigl[\tanh(\chi) (-a_n\bar a_{n} 
+ a_{n}^{*}\bar a_{n}^{*}) +(\cosh(\chi))^{-1}(a_{n}^{*}a_n + \bar a_{n}^{*}\bar a_n)\Bigr]\, .
\ee
The inverse operator has the symbol obtained by changing $\chi$ to $-\chi$ in $\tilde U_n$. 

Noting the relations
\be
C = \tanh(\chi) \, , \qquad S=\frac{1}{\cosh(\chi)} 
\ee
and comparing (\ref{Un}) with (\ref{I}) 
we see that up to a divergent determinant the relation between a (formal) operator $U=\prod_{n=0}^{\infty}U_n$ implementing the Bogolyubov 
transformation (\ref{Bog1}), (\ref{Bog2}) and the radius changing interface operator  is 
\be\label{OU}
{\cal I}^{(R_1\leftarrow R_2)} = O U^{-1} 
\ee
where $O: {\cal H}_{R_2}\to {\cal H}_{R_1}$ is a linear operator defined so that 
\be
O\, a_{n_1}^{\dagger} a_{n_2}^{\dagger} \dots |M,N\rangle_{R_2} =   b_{n_1}^{\dagger} b_{n_2}^{\dagger} \dots |M,N\rangle_{R_1} \, .
\ee
The determinant diverges because the complete canonical transformation (operating on all modes) is improper. The new vacuum has infinite norm.  
However we can define a renormalised determinant of the relevant operator $\Phi\Phi^{\dagger}$ so that  
$({\rm det}(\Phi \Phi^{\dagger})^{-1/4}$ equals 
the $g$-factor $g_{\vartheta}$ present in ${\cal I}^{(R_1\leftarrow R_2)}$. Adding the zero mode contribution and using a heat kernel type regularisation we 
can write the regularised determinant as 
\be
{\rm det}_{\epsilon}(\Phi \Phi^{\dagger}) = \exp\left( \ln(\cosh(\chi)(2 + 4\sum_{n=1}^{\infty}e^{-\epsilon n})  \right) 
= \exp\left( \ln(\cosh(\chi)(2 + 4\frac{e^{-\epsilon}}{1-e^{-\epsilon}} )\right) 
\ee
Taking $\epsilon$ to zero and subtracting the $1/\epsilon$ divergence in the exponent we obtain a renormalised value 
\be
{\rm det}_{\rm ren}(\Phi \Phi^{\dagger}) = (\cosh(\chi))^{-2}
\ee
so that 
\be
({\rm det}_{\rm ren}(\Phi\Phi^{\dagger}))^{-1/4} = \sqrt{\cosh(\chi)} = \frac{1}{\sqrt{S}} = g_{\vartheta} \, .
\ee
The subtracted operator $U$ although not being unitary in ${\cal H}_{R_2}$ satisfies the commutation relations 
(\ref{UUU}) specifying the (improper) Bogolyubov transformation. 

From the point of view of boundary conformal field theory the value of the $g$-factor for a conformal boundary 
condition $|B\rangle\!\rangle$ is fixed by Cardy constraint \cite{Cardy}
\be\label{Card}
\langle\!\langle B| e^{-2\pi H_{\rm cl}\epsilon}  |B\rangle\!\rangle = {\rm Tr} \, e^{-H_{\rm op}/\epsilon} \,  
\ee
where $H_{\rm cl}$ and $H_{\rm op}$ are the Hamiltonians corresponding to the $\tau$- and $\sigma$-quantisations respectively. 
For an interface of the kind (\ref{OU}) realising a Bogolyubov transformation, condition (\ref{Card}) is equivalent to requiring  that 
the {\it subtracted} overlap of the new vacuum 
$|0'\rangle = U|0\rangle$ 
with itself is equal to one,  which is a natural normalisation condition. 
The overlap with the old vacuum $\langle 0 |0'\rangle$ then gives the value of the $g$-factor.


\subsection{Fusion of ${\cal I}^{(R_1\leftarrow R_2)}$ with ${\cal I}^{(R_2\leftarrow R_3)}$}\label{fusion_of_defects}
The fusion of two matching interfaces: ${\cal I}^{(R_1\leftarrow R_2)}$ and ${\cal I}^{(R_2\leftarrow R_3)}$,  
is obtained by placing the two interfaces on a cylinder separated by Euclidean distance $\epsilon$ and   taking the subtracted limit
\be
{\cal I}^{(R_1\leftarrow R_2)}\circ {\cal I}^{(R_2 \leftarrow R_3)} = \lim_{\epsilon\to 0 } 
e^{d/\epsilon} {\cal I}^{(R_1\leftarrow R_2)} e^{-\epsilon H_2} {\cal I}^{(R_2\leftarrow R_3)} 
\ee
where $H_2$ is the Hamiltonian for the free boson with radius $R_2$ and $d/\epsilon$ is 
a Casimir energy counterterm. It can be represented by a picture 
\begin{center}
\begin{tikzpicture}[>=latex]
\draw[thick, dashed] (0,0) arc (270:90:0.6 and 1.1);
\draw[thick, dashed] (0,0) arc (-90:90:0.6 and 1.1 );
\draw[ thick] (0,0)--(9,0);
\draw[thick] (0,2.2 cm )--(9, 2.2 cm);
\draw[thick, dashed] (9,0) arc (270:90:0.6 and 1.1);
\draw[thick, dashed] (9,0) arc (-90:90:0.6 and 1.1);
\draw[very thick] (3.5,0) arc (270:90:0.6 and 1.1);
\draw[very thick, dashed] (3.5,0) arc (-90:90:0.6 and 1.1);
\draw[very thick] (5.8,0) arc (270:90:0.6 and 1.1);
\draw[very thick, dashed] (5.8,0) arc (-90:90:0.6 and 1.1 );
\draw (3.9,-0.4) node {${\cal I}^{(R_1\leftarrow R_2)}$};
\draw (6.4,-0.4) node {${\cal I}^{(R_2\leftarrow R_3)}$};
\draw (1.5,1.1  ) node {$R_1$};
\draw (4.6,1.1 ) node {$R_2$};
\draw (7.5,1.1 cm) node {$R_3$};
\draw (3.5, 2.25) -- (3.5,2.6);
\draw (5.8, 2.25) -- (5.8,2.6);
\draw[<->] (3.5, 2.45) -- (5.8,2.45);
\draw (4.6,2.64) node {$\epsilon$};
\end{tikzpicture}
\vspace{6mm}

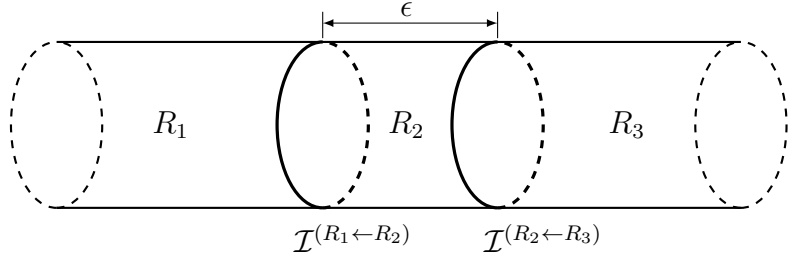
{\normalsize {\bf Figure 3}: Fusion of two radius changing interfaces.}
\end{center}


 It was found in \cite{BB} that 
\be
d = \frac{1}{2}\int\limits_{0}^{1}\frac{dx}{x}\ln(1+ CC'x) = -\frac{1}{2}{\rm Li}_{2}(-CC') 
\ee
and that 
\be\label{fusion_defs}
{\cal I}^{(R_1\leftarrow R_2)}\circ {\cal I}^{(R_2\leftarrow R_3)}= {\cal I}^{(R_1\leftarrow R_3)}\, .
\ee
The interface ${\cal I}^{(R\leftarrow R)}$ is the identity operator. The set  
${\cal G} =\{ {\cal I}^{(R_1\leftarrow R_2)} |0<R_1,R_2 \}$ thus
 forms a groupoid  
with respect to the fusion operation (\ref{fusion_defs}) with the identity element ${\cal I}^{(R\leftarrow R)}$
 and  the inverse defined as 
\be
({\cal I}^{(R_1\leftarrow R_2)})^{-1} = {\cal I}^{(R_2 \leftarrow R_1)}  \, .
\ee
We have a groupoid due to the fact that we can only fuse the interfaces with matching target and source spaces ${\cal H}_{R}$. 

The interface operators (\ref{I})  however depend essentially only on the ratio of the radii $R_{1}/R_2$ so that we can also 
associate with them a group whose elements are equivalence classes with respect to the relation: 
${\cal I}^{(R_1 \leftarrow R_2)}\sim {\cal I}^{(R_1' \leftarrow R_2')}$ if $R_1/R_2 = R_1'/R_2'$. This group is isomorphic to ${\mathbb R}^{1}$. It is 
particularly easy to see this  using the hyperbolic angles ${\chi}$ parameterising the Bogolyubov transformations to label the equivalence classes.
While the composition rule for  the angles $\vartheta$ given by equation 
\be
\tan(\vartheta'') = \frac{R_1}{R_3} = \tan(\vartheta)\tan(\vartheta') \, , \quad \tan(\vartheta) = \frac{R_2}{R_1}\, , \enspace  \tan(\vartheta')= \frac{R_3}{R_2}
\ee
is  complicated, 
the  hyperbolic 
angles 
satisfy simple addition rule: 
\be
\chi'' = \chi + \chi' \, . 
\ee


\subsection{${\cal I}^{(R_1 \leftarrow R_2)}$ as a perturbation interface} 
Here we will show that ${\cal I}^{(R_1 \leftarrow R_2)}$ can be obtained by starting with the theory with periodicity $R_2$ and 
perturbing it by the local  radius changing operator
\be
:\! \partial \phi \bar \partial \phi\!: = \frac{1}{4} [  :\! (\partial_{\tau} \phi)^{2}\!: + :\! (\partial_{\sigma} \phi)^{2}\!: ] \, .
\ee
The Euclidean action functional 
changes by 
\be \label{deltaS}
\Delta S = \lambda \int\!\! d\sigma\! \! \int\!\! d\tau :\! \partial \phi \bar \partial \phi\!: \, . 
\ee
The precise connection between the coupling $\lambda$ and the ration $R/R'$ depends on how we renormalise the perturbation theory 
divergences. One particular scheme emerges naturally when diagonalising the perturbed Hamiltonian on a cylinder. 

 In the $\tau$-quantisation 
on the Euclidean cylinder configurations on the circle $\tau=0$  give  the canonical representation of the radial quantisation Hilbert space. 
The perturbed Hamiltonian corresponding to  (\ref{deltaS}) reads
\be
H' = \frac{1}{2}(p_{L}^{2} + p_{R}^{2}) + \sum_{n=1}^{\infty} n (a_{n}^{\dagger} a_{n} + \bar a_{n}^{\dagger} \bar a_n) -\lambda \int\limits_{0}^{2\pi} d\sigma
:\! \partial \phi \bar \partial \phi\!: -\frac{1}{12}\, . 
\ee
Substituting the mode expansion (\ref{modes}) and integrating over $\sigma$ we obtain 
\be
H' =  \frac{1}{2}(p_{L}^{2} + p_{R}^{2})  +2\pi \lambda p_{L} p_{R} + 
\sum_{n=1}^{\infty} n \Bigl( a_{n}^{\dagger} a_{n} + \bar a_{n}^{\dagger} \bar a_n   + 2\pi \lambda (a_n \bar a_n + a_{n}^{\dagger} 
\bar a_{n}^{\dagger} )
\Bigr) -\frac{1}{12}\, .
\ee

The perturbed Hamiltonian $H'$ is diagonalised by a Bogolyubov transformation (\ref{Bog1}), (\ref{Bog2}) 
for which 
\be
\lambda = -\frac{1}{2\pi} \tanh(2\chi)   
\ee
so that 
\be\label{H'}
H' = \frac{1}{\cosh(2\chi)}\Bigl(  \frac{(p_{L}')^{2}}{2} + \frac{(p_{R}')^{2}}{2} + 
\sum_{n=1}^{\infty} n [ (a_{n}')^{\dagger} a_{n}' + (\bar a_{n}')^{\dagger} \bar a_n'   ]\Bigr) -\frac{1}{12}+ \Delta {\cal E}_{0}
\ee
where 
\be
\Delta {\cal E}_{0} = -\frac{\sinh^{2}(\chi)}{\cosh(2\chi)} \sum_{n=1}^{N} n
\ee
is the divergent vacuum energy shift. Here we regularised it by truncating the mode expansion at $n=N$. 
This kind of regularisation is natural for a truncated conformal space approach (TCSA) of \cite{TCSA}. 
The overall factor $(\cosh(2\chi))^{-1}$ in (\ref{H'}) gives energy scale renormalisation  discussed in the context of TCSA
in \cite{TCSA_Watts}, \cite{TCSA_Rychkov}. The energy rescaling was to be expected as the perturbation shifts the kinetic term in  action. 

   From (\ref{Bog2}) we can express the coupling constant via the radii 
   \be\label{lambda1}
   \lambda = -\frac{1}{2\pi} \left(\frac{R_1^4 - R_2^4}{R_1^4 + R_2^4}\right) \, .
   \ee
   We note that this expression is different from the correspondence worked out in conformal perturbation theory 
   in \cite{Moore}, \cite{Melzer} where 
   \be
   \lambda_{\rm CP} = -\frac{1}{\pi} \tanh(\chi) = -\frac{1}{\pi} \left(\frac{R_1^2 - R_2^2}{R_1^2 + R_2^2}\right) \, .
      \ee
   The two schemes differ by a coupling constant reparameterisation. We note that the TCSA regularisation scheme breaks Lorentz 
   invariance  and quantities computed in it may be different from the ones obtained using a Lorentz invariant  regulator 
   \cite{TCSA_Konik}. 
   

\subsection{${\cal I}^{(R_1 \leftarrow R_2)}$ and  transport of states} 

A change of radius for a free compact boson is an example of a marginal deformation of a conformal field theory. 
In operator formalism (see e.g. \cite{Vafa} or section 2 of \cite{CNW}) a CFT is described in terms of surface states. 
Let $\Sigma$ be a Riemann surface with punctures $p_1, \dots, p_n$ and local coordinates $z_{1}, \dots , z_n$ that vanish 
at the respective punctures. A CFT assigns to every such surface a surface state  in an n-fold tensor product of 
the state space ${\cal H}$:  $|\Sigma; z_1, \dots , z_n\rangle \in {\cal H}\otimes \dots \otimes {\cal H} $.
This state can be thought of being obtained by performing a functional integral over $\Sigma$ minus parameterised 
circles around the puncture at each of which we have a copy of ${\cal H}$ defined in configuration space.

A deformation of a given CFT can be described  in terms of deformed surface states (see \cite{CNW} and references within). 
The surface states can be deformed by integrating the deformation operator over the Riemann surface minus identical 
disks cut around the punctures. 
At the leading order the change in the surface states is 
\be \label{infinites_def}
\delta |\Sigma;z_1,\dots , z_n\rangle = \int\limits_{\Sigma-\cup_{i}D_i}d^2z\, \langle \phi(z)| \Sigma; z_1,\dots z_n, z\rangle  
\ee
where the bra state $\langle \phi(z)|$ corresponds to the deformation operator $\phi$ being inserted at $z$, 
the surface state  $| \Sigma; z_1,\dots z_n, z\rangle$ corresponds to the original surface $\Sigma$ with punctures at $z_i$ and 
with an additional puncture at $z=0$, and the integration is taken over $\Sigma$ minus the unit discs cut out in the $z_i$ 
coordinates around the punctures $p_i$.   

Associated with this deformation formula is 
 a canonical flat connection $\hat \Gamma$ on the deformation moduli space \cite{Ranganathan}, \cite{RSZ} which 
 can be used to construct parallel transport of states between the undeformed and deformed state spaces. 
 Formula (\ref{infinites_def}) for the infinitesimal deformation generalises naturally to a finite deformation via standard 
 perturbative expansion of $\exp( \int d^2 z \phi(z)) $. It is clear that what we obtain is a collection of perturbation interfaces
 placed on circles around the punctures $p_i$. In constructing the interfaces multiple collisions with the boundaries of the disks 
 are regulated and divergences are subtracted. A second order linear divergence arising in integrating the connection $\hat \Gamma$ 
 was noted in section 7.1 of \cite{RSZ}. We note that it is the same as the leading order linear  divergence in the $g$-factor of perturbation 
 interfaces (defects)  discussed in \cite{KSC}, it is associated with the boundary identity field that lives on the boundary of the integration region.   
 

It was shown in \cite{Ranganathan} that the Bogolyubov transformation (\ref{Bog1}), (\ref{Bog2}) 
is infinitesimally equivalent to the transport associated with the connection $\hat \Gamma$ for the radius changing  deformation of a free boson. 
Since we showed that (\ref{Bog1}) and (\ref{Bog2}) is realised by the interface ${\cal I}^{(R_1 \leftarrow R_2)}$ it 
follows, assuming that the infinities associated with integrating $\hat \Gamma$ 
are subtracted in accordance with the finite version of the Bogolyubov transformation (\ref{Bog1}), 
(\ref{Bog2}),  that this interface realises the (finite) 
transport associated with connection $\hat \Gamma$. As is clear from (\ref{I}) any  state in ${\cal H}_{R_2}$ with finitely many particles (oscillators) 
in mapped into into a state in ${\cal H}_{R_1}$  with infinitely many particles. In particular the vacuum is mapped into a squeezed state. 
In CFT language each primary is  mapped into an infinite combination of descendants. 


\section{Fusion with $D$-branes. }\label{section_fusion1}
\setcounter{equation}{0}

As discussed in the introduction the conformal interface  ${\cal I}^{(R_1\leftarrow R_2)}$ can be fused with a conformal boundary state $|B\rangle\! \rangle_{R_2}$ 
using the subtracted limit
\be
|{\cal I}^{(R_1\leftarrow R_2)} \circ B\rangle\!\rangle_{R_1} = \lim_{\epsilon \to 0} e^{d/\epsilon} {\cal I}^{(R_1 \leftarrow R_2 )}e^{-\epsilon H_2} |B\rangle\!\rangle_{R_2} \, .
\ee

As a warmup we will work out in detail the  fusion of  ${\cal I}^{(R_1\leftarrow R_2)}$ with Dirichlet and Neumann branes. We will 
see that it  gives again the Dirichlet and Neumann branes 
respectively at the new radius. 
The Dirichlet and Neumann boundary states are 
\be\label{Dir}
|\!|D\rangle\!\rangle_{R_2} = \frac{1}{\sqrt{R_2}}  \prod_{n=1}^{\infty} \exp(a_{n}^{\dagger}\bar a_n^{\dagger}) \sum_{N=-\infty}^{\infty} 
e^{-2iN\psi_{0}/R_{2}} |N,0\rangle_{2} \, , 
\ee
\be\label{Neum}
|\!|N\rangle\!\rangle_{R_2} = \sqrt{\frac{R_2}{2}} \prod_{n=1}^{\infty} \exp(-a_{n}^{\dagger}\bar a_n^{\dagger})  \sum_{M=-\infty}^{\infty} 
e^{iM\tilde \psi_{0}R_2} |0,M\rangle_{2}\, . 
\ee
Here $\psi_0$ and $\tilde \psi_{0}$ are the position and Wilson line moduli. 

For the fusion of the Dirichlet brane with ${\cal I}_{\vartheta}$ we find using (\ref{I})
\be
{\cal I}^{(R_1 \leftarrow R_2)} q^{L_0^{(2)} + \bar L_0^{(2)}}  |\!|D\rangle\!\rangle =
\frac{g_{\vartheta}q^{-\frac{1}{12}}}{\sqrt{R_2}}\sum_{N=-\infty}^{\infty} e^{-2iN\psi_{0}/R_{2}} q^{N^2/R_{2}^2} \prod_{n=1}^{\infty}e^{-Cb_n^{\dagger}\bar b_{n}^{\dagger}} 
\hat {\cal A}_{n}|N,0\rangle_1 
\ee
where $q=e^{-\epsilon}$ and 
\be
\hat {\cal A}_{n} = \langle 0| \exp\Bigl[ Ca_{n}\bar a_{n} 
+S(b_{n}^{\dagger}a_{n}+ \bar b_{n}^{\dagger}\bar a_{n})   \Bigr] e^{q^{2n}a_{n}^{\dagger} \bar a_{n}^{\dagger}}|0\rangle\,  
\ee
are operators in ${\cal H}_{R_1}$. 
Using integral representations 
\be\label{int1}
e^{Ca_n \bar a_{n}} = \int \!\frac{d^2 z}{\pi} e^{-z\bar z -zCa_n -\bar z\bar a_n}\, , 
\ee
\be\label{int2}
e^{q^{2n}a_n^{\dagger} \bar a_{n}^{\dagger}} = \int\! \frac{d^2 w}{\pi}e^{-w\bar w - wq^{n}a_{n}^{\dagger} - \bar w\bar a_{n}^{\dagger}q^n}
\ee
we obtain 
\bea
&& \hat {\cal A}_{n} =  \int\! \frac{d^2 z}{\pi} \int\! \frac{d^2 w}{\pi}e^{-z\bar z - w\bar w} 
\exp\Bigl[  (-zC+Sb_{n}^{\dagger})(-wq^n) + (-\bar z + S\bar b_{n}^{\dagger})(-\bar w q^n) \Bigr] \nonumber
 \\
&&= \frac{1}{1-Cq^{2n}} \exp\Bigl[ \frac{q^{2n}S^2b_{n}^{\dagger}\bar b_{n}^{\dagger}}{1-Cq^{2n}} \Bigr]\, . 
\eea
Assuming $C\ne 1$ (that is $R_1\ne \infty$) we can extract the leading divergence using  Euler-Maclaurin formula 
\bea \label{DCas1}
&& \prod_{n=1}^{\infty} \frac{1}{1-Cq^{2n}} = \exp\Bigl[ -\sum_{n=1}^{\infty} \ln(1-Ce^{-2n\epsilon}) \Bigr]
\nonumber \\
&& = \exp\Bigl[ - \frac{1}{2\epsilon}\int\limits_{0}^{1}\! \frac{dx}{x}\ln(1-Cx) + \frac{1}{2}\ln(1-C) + \frac{\epsilon C}{6(1-C)} +  {\cal O}(\epsilon^2)\Bigr]\, .
\eea
The  Casimir energy is thus 
\be \label{DCas2}
{\cal E}_{0}={\cal E}_{D}\equiv \frac{1}{2}\int\limits_{0}^{1}\! \frac{dx}{x}\ln(1-Cx)= -\frac{1}{2}{\rm Li}_{2}(C)
\ee
We note that for $C>0$ (increasing the  radius) ${\cal E}_{D}$  is negative so that the fusion amplitude diverges while for 
$C<0$ (decreasing the radius) ${\cal E}_{D}>0$ so that the fusion amplitude goes to zero.  
The term $\frac{1}{2}\ln(1-C) $ in the exponent is a shift of boundary entropy which corrects the $g$-factor of 
$|D\rangle\!\rangle_{R_2}$ into that of $|D\rangle\!\rangle_{R_1}$. Thus 
\be
|{\cal I}^{(R_1 \leftarrow R_2)}\circ D\rangle\!\rangle_{R_2} = |D\rangle\!\rangle_{R_1}
\ee
where the position modulus $\psi_{0}$ is rescaled by a factor $R_2/R_1$. 

For the fusion with the Neumann brane the analogous calculation gives 
\be
|{\cal I}^{(R_1 \leftarrow R_2)}\circ N\rangle\!\rangle_{R_2} = |N\rangle\!\rangle_{R_1}
\ee
with the Wilson line  modulus $\tilde \psi_{0}$ rescaled by a factor $R_1/R_2$. 
The Casimir energy of the fusion is 
\be\label{dN}
{\cal E}_{0} = {\cal E}_{N}\equiv  \frac{1}{2}\int\limits_{0}^{1}\! \frac{dx}{x}\ln(1+Cx)= -\frac{1}{2}{\rm Li}_{2}(-C)
\ee
which has the opposite sign to ${\cal E}_{D}$  so that when the radius increases the  fusion amplitude goes to zero.

For  generic radius the only D-branes with finite $g$-factor are the Dirichlet and Neumann branes and 
their superpositions \cite{Friedan}. Changing the radius transports these $D$-branes to  superpositions of the same kind.
Looking at it perturbatively, the bulk perturbation corresponding to changing the radius does not trigger an RG flow 
on the boundary. At the radii given by $R=\frac{p}{q}R_{\rm s.d.}$ where $p$ and $q$ are integers and 
 $R_{\rm s.d.}=\sqrt{2}$ is the self-dual radius there are exceptional D-branes. 
In particular at the self dual radius $R=R_{\rm s.d.}$ the irreducible D-branes are parameterised by a group element $ g\in {\rm SU}(2)$. 
We denote the corresponding boundary states as $|g\rangle\!\rangle$. 
As argued in \cite{Gaberdiel_etal} changing the radius for generic $g$ (that does 
not correspond to Neumann or Dirichlet branes) induces a boundary RG flow. The end point of the flow is a Dirichlet brane 
if the new radius is larger and a Neumann brane if the new radius is smaller. 


As discussed in section \ref{fusion_of_defects}  the interfaces ${\cal I}^{(R' \leftarrow R)}$ form a groupoid ${\cal G}$ with respect to the fusion operation. 
It may seem plausible  that fusion with D-branes defines an action of ${\cal G}$ on the space of conformal boundary conditions. 
This would mean that the following rule holds 
\bea\label{gr_action}
 {\cal I}^{(R_1 \leftarrow R_2)}\circ ({\cal I}^{(R_2 \leftarrow R_3)}\circ |B\rangle\!\rangle_{R_3}) && = 
({\cal I}^{(R_1 \leftarrow R_2)}\circ {\cal I}^{(R_2 \leftarrow R_3)})\circ |B\rangle\!\rangle_{R_3} \nonumber \\
&&  ={\cal I}^{(R_1 \leftarrow R_3)}\circ |B\rangle\!\rangle_{R_3}\,  
\eea
for all boundary states $|B\rangle\!\rangle_{R_3}$.
This rule does hold for the Dirichlet and Nuemann branes but already for their superposition it breaks down. 
Thus, using (\ref{DCas2}), (\ref{dN}) we find  
\be
{\cal I}^{(R_1\leftarrow R_2)} \circ (|D\rangle\!\rangle_{R_2} + |N\rangle\!\rangle_{R_2}) = 
\left \{
\begin{array}{l@{\qquad}l}
|D\rangle\!\rangle_{R_1}\, ,  & \mbox{ if } R_1>R_2 
\\[1ex]
|N\rangle\!\rangle_{R_1}\, ,  & \mbox{ if } R_1<R_2 
\end{array}
\right .
\ee
so that we cannot get back to the original boundary condition if we apply the inverse interface ${\cal I}^{(R_2\leftarrow R_1)}$.
This has the following RG flow interpretation. When we perturb the bulk theory by the radius changing operator the only 
relevant operator that appears in the bulk-to-boundary OPE in each of the two boundary components: $|D\rangle\!\rangle_{R_2}$, 
$|N\rangle\!\rangle_{R_2}$,
is the corresponding component of identity: ${\bf 1}_{\rm D}$, ${\bf 1}_{ \rm N}$. If we use the minimal subtraction scheme 
to remove the associated power divergences then the RG flow would leave each term in the superposition intact. 
If however we add finite counterterms then the couplings corresponding to ${\bf 1}_{ \rm D}$ and ${\bf 1}_{ \rm N}$ flow  
 with the RG time which in general will result in one component exponentially dominating over the other. The Casimir 
 energies (\ref{DCas2}), (\ref{dN}) correspond to a particular non-minimal RG scheme associated with ${\cal I}^{(R_1\leftarrow R_2)}$ 
 in which the identity component couplings flow. This is an example of nontrivial functions $\beta_{(0)}^{i}(\lambda)$, 
(for $i$ corresponding to the fields ${\bf 1}_{ \rm D}$, ${\bf 1}_{ \rm N}$)
 discussed in the introduction. 

 In general we believe that whenever the fusion of an interface with a boundary condition induces a non-trivial boundary RG flow  
 the result of the fusion won't be invertible. In particular we expect that the fusion ${\cal I}^{(R \leftarrow \sqrt{2})}\circ |g \rangle\!\rangle$ cannot be inverted for a generic $g$. 



\subsection{Fusion with exceptional branes at the self-dual radius}
From now on we take $R_2=\sqrt{2}$, $R_1\equiv R$  and study the fusion of ${\cal I}^{(R \leftarrow \sqrt{2})}$ with the
 exceptional branes $|g\rangle\!\rangle $ in more detail. At the self-dual radius the symmetry algebra is enlarged to ${\rm su}(2)_{1}\oplus {\rm su}(2)_{1}$ 
 current algebra with holomorphic generators 
 \be\label{su2_currents}
 J^{3}(z) =  \frac{i}{\sqrt{2}}\partial \phi(z)\, , \quad   J^{+}(z) = :\! e^{i\sqrt{2}\phi(z)}\!: \, ,  \quad 
 J^{-}(z) = :\! e^{-i\sqrt{2}\phi(z)}\!: 
  \ee
 and similar expressions for the antiholomorphic ones. The zero modes $J^{3}_{0}$, $J^{\pm}_{0}$  of the holomorphic currents generate 
 the ${\rm su}(2)$ Lie algebra. The exponents of the generators give a representation of ${\rm SU}(2)$ group. 
 For $g\in  {\rm SU}(2)$ we 
 will denote the corresponding operators acting in ${\cal H}_{\sqrt{2}}$ by $\hat g$.
 
 We will focus on the exceptional branes  with the boundary state
\be
|g\rangle\!\rangle = \hat g |D\rangle\!\rangle_{\sqrt{2}}
\ee 
where for simplicity we will take $\psi_{0}=0$ and assume that 
\be \label{ghat}
\hat g = e^{\pi i \mu (J^{+}_{0} + J^{-}_{0})} 
\ee
that corresponds to a group element 
\be \label{g}
g\equiv g(\mu)=\left( \begin{array}{rr} \cos(\pi \mu) & i\sin(\pi \mu)\\
i\sin(\pi \mu) & \cos(\pi \mu)
\end{array}  
\right) \, .
\ee
Here $\mu$ is a real parameter that runs from $\mu=0$ that corresponds to a Dirichlet brane, to 
$\mu=1/2$ that corresponds to a Neumann brane. This follows from the fact that the group element 
\be \label{T}
 T  \equiv \left( \begin{array}{rr} 0 & 1\\
1 & 0
\end{array}  
\right) 
\ee
realises the T-duality action via the corresponding $\hat T$ action. 
Note that 
\be
g(\mu) T = i g\left(\mu - \frac{1}{2}\right)  
\ee
that gives the T-duality action on the boundary conditions labelled by $\mu$.

Most of our results below can be easily generalised to the general case with 
\be
\hat g = e^{\pi i (\mu J^{+}_{0} + \bar \mu J^{-}_{0})} e^{\pi i \nu J_{3}}  
\ee
where $\mu$ is complex and $\nu$ is real. At $\mu=0$ the value of $\nu$ gives the $\psi_0$ modulus while 
at $|\mu|=1/2$, $\nu=0$ we have a Neumann brane with the phase of $\mu$ specifying $\tilde \psi_0$.

 To determine the fusion brane $|{\cal I}^{(R\leftarrow \sqrt{2})}\circ g\rangle\!\rangle_{R}$ it is enough to determine its Ishibashi states 
 content. To this end it is enough to study the overlaps 
 \be
 {}_{R}\langle X| {\cal I}^{(R \leftarrow \sqrt{2})} q^{L_{0}^{(2)} + \bar L^{(2)}_{0}} \hat g |D\rangle\!\rangle_{\sqrt{2}}
  \ee 
 where $|X\rangle_{R}  $ are Virasoro primaries of zero spin.
Assuming $R$ is not a rational of the self-dual radius the only spinless primary states with nonzero momentum 
 are the states $|N,0\rangle_{R} $, $|0,M\rangle_{R}$. At zero momentum there are additional primaries 
 of dimension $h_{n} = \frac{1}{2}n^2$, $n\in {\mathbb Z}$ which are $U(1)$-descendants of the vacuum. 
 We will focus on the non-zero momentum primaries and the vacuum. 

It will be instructive to start with an arbitrary $U(1)$ primary and see how spin conservation works. 
It will be convenient to label the $U(1)$ primaries in ${\cal H}_{\sqrt{2}}$ by their left and right momentum 
${}_{\sqrt{2}}\langle N, M| \equiv \langle p_L, p_R| $
so that $|N,0\rangle_{\sqrt{2}} \equiv |N/\sqrt{2}, N/\sqrt{2}\rangle $. Thus we consider the amplitudes
\bea \label{ampl1}
&& \langle p_L, p_{R}| {\cal I}^{(R \leftarrow \sqrt{2})} q^{L_{0}^{(2)} + \bar L^{(2)}_{0}} \hat g |D\rangle\!\rangle_{\sqrt{2}}
  \nonumber \\
 && = 2^{-\frac{1}{4}}  \langle p_{L},p_{R}| \prod_{n=1}^{\infty} \exp\Bigl[ C a_{n}\bar a_{n} 
   \Bigr]  q^{L_0^{(2)} + \bar L_0^{(2)}} \hat g \prod_{m=1}^{\infty} \exp(a_{m}^{\dagger}\bar a_m^{\dagger}) \sum_{L\in {\mathbb Z}} 
 |\frac{L}{\sqrt{2}}, \frac{L}{\sqrt{2}} \rangle \, 
 \eea
 where 
 \be
 C=\frac{R^2-2}{R^2 + 2} \, .
 \ee
 Performing the contractions in the anti-holomorphic sector (that commutes with $\hat g$) we 
rewrite (\ref{ampl1}) as an amplitude in a tensor product of two holomorphic Fock spaces 
\be\label{ampl_red}
 2^{-\frac{1}{4}} \langle p_{L}| 
\sum_{k_1, k_2, \dots } C^{k_1 + k_2 + \dots} q^{2(k_1 + 2k_2 + 3k_3 + \dots)} 
\frac{a_1^{k_1}a_2^{k_2} \dots \hat g (a_1^{\dagger})^{k_1} (a_2^{\dagger})^{k_2} \dots }{k_1 !  k_2 ! k_3 ! \cdot \dots } |p_{R}\rangle \, .
\ee

To show  that (\ref{ampl_red}) is non-vanishing only if $p_{L}=\pm p_{R}$.  
 we rewrite it in terms of  the $\widehat{su}(2)$ current modes. We have 
\be
a_{n}=\sqrt{\frac{2}{n}}J^{3}_{n} \,  , \quad a^{\dagger}_{n} = \sqrt{\frac{2}{n}}  J^{3}_{-n}   \, , \quad n>0 \, 
\ee
and similarly  for $\bar a_n$, $\bar a_n^{\dagger}$.
The $\widehat{\rm su}(2)$ algebra is 
\bea \label{su2cr1}
&&[J_{n}^{3}, J_{m}^{3}]=\frac{n}{2}\delta_{n+m,0} \, , \qquad [J_{n}^{+}, J_{m}^{-}]=2J^{3}_{n+m} + n\delta_{n+m,0} \\ \label{su2cr2}
&& [J_{n}^{3},J^{+}_{m}] = J^{+}_{n+m} \, , \qquad  [J_{n}^{3}, J_{m}^{-}]=-J^{-}_{n+m} \, .
\eea

We can move  $\hat g$ through all of the creation operators to the right  in (\ref{ampl_red}) using 
\be \label{rotated_J}
J_{-n}(\mu) \equiv \hat g J^{3}_{-n} \hat g^{-1} = \cos(2\pi \mu)J_{-n}^{3} + \frac{1}{2i}\sin(2\pi \mu) (J_{-n}^{+}-J^{-}_{-n}) \, . 
\ee
It follows then from (\ref{ghat}) and the commutation relations (\ref{su2cr2})  that 
(\ref{ampl_red}) can be represented as a sum of  amplitudes of the form
\be
\langle p_{L}| J^{\epsilon_1}_{m_1}J^{\epsilon_2}_{m_2} \dots J^{\epsilon_k}_{m_k}|p_R\rangle 
\ee 
where each $\epsilon_i=\pm $ and the $m_j$ are integers such that their sum equals to zero: $\sum m_j = 0$.
This means that the state 
$$
J^{\epsilon_1}_{m_1}J^{\epsilon_2}_{m_2} \dots J^{\epsilon_k}_{m_k}|p_R\rangle 
$$
has weight $\frac{1}{2}p_{R}^2$ and charge $p_R + \sum_i \epsilon_i $. Unless $p_{R} + \sum_i \epsilon_i = \pm p_{R}$ 
this state is a U(1) descendant and thus its inner product with $\langle p_{L}|$ vanishes.
(For the given weight there are only two U(1) primaries.) Note that the amplitude (\ref{ampl_red}) for $p_L=-p_R$ can be expressed in terms of the one with $p_L=p_R$ 
with $C$ changed to $-C$ and $\hat g$ changed by multiplying it by the T-duality operator $\hat T$.

To summarise we have obtained 
\be
 \langle p_L, p_{R}| {\cal I}^{(R_1\leftarrow \sqrt{2})} q^{L_{0}^{(2)} + \bar L^{(2)}_{0}} \hat g |D\rangle\!\rangle_{\sqrt{2}} = 2^{-\frac{1}{4}}[ \delta_{p_{L}, p_{R}} 
 T_{p_L}(\hat g ; q,C) + \delta_{p_{L}, -p_{R}} 
 \tilde T_{p_L}(\hat g;q,C)] \, 
  \ee
where 
\be
\tilde T_{p}(\hat g;q,C) = T_{p}(\hat g\hat T ; q,- C)
\ee
and 
the basic amplitude 
\be\label{basic_amp}
T_{p}(\hat g; q,C) = {\rm Tr}(  \Pi_{p}q^{2L_0} C^{N} \hat g)
\ee
is defined so that the trace is taken over the free boson representation space\footnote{Depending on the value of $p$ we can 
restrict the trace to one of the two irreducible representations of ${\widehat{\rm su}(1)}_1$.}  of ${\widehat{\rm su}(1)}_1$ 
\be\label{Hchiral}
{\cal H}^{\rm chiral} = \bigoplus_{N\in {\mathbb Z}} {\cal F}_{N/\sqrt{2}} \, , 
\ee
$\Pi_{p}$ is the orthogonal projector on the chiral Fock subspace ${\cal F}_{p}$ with momentum $p$, 
and $N$ is the oscillator number operator 
\be
N (a_1^{\dagger})^{k_1} (a_2^{\dagger})^{k_2} \dots  |p \rangle = (k_1 + k_2 + \dots ) (a_1^{\dagger})^{k_1} (a_2^{\dagger})^{k_2} \dots  |p \rangle\, . 
\ee
The amplitude $T_{p_L}(\hat g; q,C)$ multiplies the Dirichlet type Ishibashi states $|N,0\rangle \!\rangle_{R}$ while 
$\tilde T_{p_L}(\hat g; q,C)$ multiplies the Neumann type ones $|0,M\rangle \!\rangle_{R}$. The $q\to 1$ leading behaviour of these amplitudes 
determines which type of the two amplitudes survives in the fusion. 

  
  So far we were not able to find a method to calculate $T_{p}(\hat g; q,C)$ to all orders in $\mu$ and $C$. 
  Algebraically, the extra weighting $C^{N}$ proves to be tricky to take into account. Although $L_{0}$ commutes with both $N$ 
   and $\hat g$, the operator $N$ does not commute with $\hat g$ unless $g$ is in the U(1) subgroup generated by $J^{3}_{0}$. 
 There is to the best of our knowledge no useful algebraic structure which includes both   $N$ and the representation $\hat g$. 
 In the next section we develop separate perturbation series in 
  $\mu$ and $C$. This allows us to observe the RG logarithms and to see how they can be resummed in the complete 
  amplitudes $T_{p}(\hat g; q,C)$. Later we  calculate the     $T_{p}(\hat g; q,1)$ amplitudes non-perturbatively and 
  find some  evidence for the conjecture that fusion with ${\cal I}^{(R\leftarrow \sqrt{2})}$ produces the end point of  the RG flow 
 of \cite{Gaberdiel_etal}. 
    
    
    
  \section{Perturbative calculations}  \label{sec_pert}
\setcounter{equation}{0}
\subsection{Perturbation series in $C$ and RG logarithms}

Define the expansions 
\be\label{Cexp}
T_{p}(\hat g, C; q)= \sum_{k=0}^{\infty} C^{k} T_{p}^{(k)}(\hat g;q) \, , \qquad 
\tilde T_{p}(\hat g, C; q)= \sum_{k=0}^{\infty} C^{k} \tilde T_{p}^{(k)}(\hat g;q) \, .
\ee
To calculate $T_{p}^{(k)}(\hat g;q)$ we first calculate the amplitudes of the form 
\be \label{ampl0}
{\cal A}_{p; n_1, \dots n_k}\equiv \langle p| J_{n_1}^{3}\dots J_{n_k}^{3} \hat g J_{-n_{k}}^{3} \dots J_{-n_1}^3 |p\rangle 
\ee
and then perform summations over $n_1, \dots , n_k$. To calculate (\ref{ampl0}) we first pull $\hat g$ through the oscillators 
to the right. Using (\ref{rotated_J}) we rewrite (\ref{ampl0}) as 
\be 
{\cal A}_{p; n_1, \dots n_k} = \langle p| J_{n_1}^{3}\dots J_{n_k}^{3}  J_{-n_{k}}(\mu) \dots J_{-n_1}(\mu) \hat g |p\rangle 
\ee
This amplitude can be calculated in the language of $\widehat{{\rm su}}(2)_{1}$ representation using the representations 
of U(1) primary states 
\bea
&&|n\sqrt{2}\rangle = J^{+}_{-2n+1}J^{+}_{-2n+3} \dots J^{+}_{-1}|0\rangle \, , \nonumber \\
&&\left|\frac{2n+1}{\sqrt{2}}\right\rangle = J^{+}_{-2n}J^{+}_{-2n+2} \dots J^{+}_{-2}\left|\frac{1}{\sqrt{2}}\right\rangle
\eea
where $n$ is a positive integer and the analogous representations for negative momenta. 

Charge zero  amplitudes are particularly easy to evaluate. We find 
\be
T_{0}^{(0)}(\hat g;q)=1\, ,
\ee
\be\label{T1}
T_{0}^{(1)}(\hat g;q)= \sum_{n=1}^{\infty}\langle 0| q^{2n}a_n \hat g a^{\dagger}_{n}|0\rangle = \cos(2\pi \mu) \frac{q^2}{1-q^2} \, ,
\ee
\bea \label{llog}
&& T_{0}^{(2)}(\hat g;q)= \sum_{n<m} \langle 0| q^{2(n+m)} a_n a_m \hat g a_{n}^{\dagger}a_{m}^{\dagger} |0\rangle 
+ \frac{1}{2!}\sum_{n=1}^{\infty}\langle 0| q^{4n} (a_n)^2 \hat g (a_n^{\dagger})^{2}|0\rangle = \nonumber \\
&& \cos^2(2\pi \mu) \frac{q^4}{(1-q^2)(1-q^4)} + \sin^2(2\pi \mu) \Bigl[ \ln\left( \frac{1-q^2}{1+q^2} \right)  
+ 2\frac{\ln(1+q^2)}{1-q^2}  \Bigr]\, .
\eea
The last expression contains a logarithmic divergence $\ln(1-q)$.  The coefficient at the divergence can be rewritten in terms 
of the matrix elements of $g$ given in (\ref{g}). Denoting as in \cite{Gaberdiel_etal}
\be
g=\left( \begin{array}{rr}
a&b^{*}\\
-b&a^{*}
\end{array}\right) 
\ee
we have  $\sin^2(2\pi \mu) = 4 |a|^2|b|^2$.    This matches with the logarithmic divergence in the perturbed 1-point function of 
$:\!\!\partial \phi\bar \partial \phi\!\!:$
found in \cite{Gaberdiel_etal} (see section 3.1 in that paper, in particular formulas (3.11), (3.12)). 

Other simple amplitudes that exhibit a logarithmic divergence  are 
$T^{(1)}_{\frac{1}{\sqrt{2}}}(\hat g; q)$ and $\tilde T_{\frac{1}{\sqrt{2}}}^{(1)} (\hat g ; q)$. 
We find using (\ref{rotated_J})
\bea
&& T^{(1)}_{\frac{1}{\sqrt{2}}}(\hat g; q)= \sum_{n=1}^{\infty} q^{2n} \left\langle \frac{1}{\sqrt{2}}\right| a_n \hat g a_{n}^{\dagger} 
\left|\frac{1}{\sqrt{2}}\right\rangle \nonumber \\
&& =\frac{q^2}{1-q^2} \cos(2\pi \mu) D_{1/2,1/2}^{1/2}(g) + i \ln(1-q^2) \sin(2\pi \mu) D_{-1/2,1/2}^{1/2}(g) \, ,
\eea
\bea
&& \tilde T_{\frac{1}{\sqrt{2}}}^{(1)} (\hat g ; q)= \sum_{n=1}^{\infty} q^{2n} \left\langle \frac{1}{\sqrt{2}}\right| a_n \hat g a_{n}^{\dagger} 
\left|-\frac{1}{\sqrt{2}}\right\rangle \nonumber \\
&& =\frac{q^2}{1-q^2} \cos(2\pi \mu) D_{1/2,-1/2}^{1/2}(g) + i \ln(1-q^2) \sin(2\pi \mu) D_{-1/2,-1/2}^{1/2}(g) \, .
\eea
Here $D^{j}_{m,n}(g)$ are matrix elements of $g$ in representation with spin $j$. Explicit formulas for them are known  
(see e.g. formula (3.4) in \cite{GRW}). For the case at hand the relevant matrix elements are given by (\ref{g}) 
  \be
 D_{1/2,1/2}^{1/2}(g) = a=\cos(\pi \mu) \, , \qquad 
 D_{-1/2,1/2}^{1/2}(g)=b^{*}=i\sin(\pi \mu) \, , 
   \ee
 \be
 D_{1/2,-1/2}^{1/2}(g) = -b=i\sin(\pi \mu) \, , \qquad 
 D_{-1/2,-1/2}^{1/2}(g)=a^{*}=\cos(\pi \mu) 
   \ee
 so that 
 \be
 T^{(1)}_{\frac{1}{\sqrt{2}}}(\hat g; q)=\frac{q^2}{1-q^2} \cos(2\pi \mu)\cos(\pi \mu) 
 -\ln(1-q^2)\sin(2\pi \mu) \sin(\pi \mu)\, , 
   \ee
 \be\label{charge_log}
 \tilde T_{\frac{1}{\sqrt{2}}}^{(1)} (\hat g ; q)= i \frac{q^2}{1-q^2} \cos(2\pi \mu)\sin(\pi \mu) 
 +i\ln(1-q^2)\sin(2\pi \mu) \cos(\pi \mu) \, . 
  \ee
 It is not hard to see that these logarithmic divergences are associated with the first order perturbation divergences in the 
 respective 1-point functions 
 \be
 \langle V_{1/2,1/2} \rangle \equiv \langle :\!\!e^{\frac{i}{\sqrt{2}}\phi}\!\!:\rangle  \, , \qquad 
  \langle V_{1/2,-1/2} \rangle \equiv \langle :\!\!e^{\frac{i}{\sqrt{2}}(\phi_L - \phi_R)}\!\!:\rangle \, .
   \ee
   To get a better picture consider the 1st order perturbation integral
   \bea
  && \int\!\! d^{2}z \langle V_{1/2,-1/2}(w, \bar w) :\!J^{3}\bar J^3\!:(z, \bar z) \rangle  = 
  \int\!\! d^2 z \langle V_{1/2,-1/2}(w, \bar w) J^{3}(z) \Bigl[J^{3}(\bar z) \cos(2\pi \mu) + \nonumber \\
  && + \frac{1}{2i}(J^{+}(\bar z) - J^{-}(\bar z))\sin(2\pi \mu) \Bigr]\rangle \, .
      \eea
      Here we integrate over a half plane ${\rm Im} z \ge 0$ with a boundary condition specified by $|g\rangle\!\rangle$.
    This integral  contains a logarithmic divergence that arises from the OPE 
    \be
    J^{3}(z) J^{-}(\bar z) \sim -\frac{J^{-}(\bar z)}{z-\bar z} \, .
    \ee  
      This divergence  is therefore of the same origin
       as the  divergence that gives raise to the leading term in the beta function (\ref{leading_beta}) 
       and comes from the bulk-to-boundary OPE. 
      We note that the logarithm in $T^{(1)}_{\frac{1}{\sqrt{2}}}(\hat g; q)$ first appears at the order $\mu^2$ while in 
$\tilde T^{(1)}_{\frac{1}{\sqrt{2}}}(\hat g; q)$ it appears at $\mu^1$.

 \subsection{The fusion singularities in the vacuum sector}\label{vacuum_sing2}
 As discussed in the introduction although we do not know what kind of singularity to expect for a general 
 amplitude  $T_{p}(\hat g; q,C)$, for the vacuum amplitude $T_{0}(\hat g; q,C)$ we expect an essential singularity 
 \be
 T_{0}(\hat g; q,C) \sim g(\mu , C) e^{-{\cal E}_{0}(\mu, C)) /\epsilon} \, , \qquad \mbox{ for } q=e^{-\epsilon} \to 1
 \ee
From $T^{(1)}_{0}$ in (\ref{T1}) we find that, assuming $g(\mu , C)$, ${\cal E}_{0}(\mu, C)$ have  perturbative expansions 
in $C$, the leading corrections are   
\be \label{fusion_pert}
{\cal E}_{0}(\mu, C) = -\frac{C}{2}\cos(2\pi \mu) + {\cal O}(C^2) \, , \qquad 
g( \mu, C) = \frac{1}{2^{1/4}}( 1 - \frac{C}{2}\cos(2\pi \mu) + {\cal O}(C^2)) 
\ee
 This means that the factor $g( \mu, C) $ has a non-trivial dependence on $\mu$ and thus the fusion of 
 ${\cal I}^{(R\leftarrow \sqrt{2})}$ with 
 $|g\rangle\!\rangle$ cannot reproduce the $g$-factor of either of the two conformal boundary conditions: $|D\rangle\!\rangle_{R}$,  $|N\rangle\!\rangle_{R}$ 
 exactly but rather will multiply it by some function that depends on $g\in {\rm SU}(2)$. This picture is also supported by our $C=1$ calculations 
 in section \ref{C1}.
 

\subsection{Perturbation series in $\mu$}
Another perturbative expansion we can develop is in $\mu$ using the free boson representation of the 
currents (\ref{su2_currents}). 
Define the expansion coefficients\footnote{Only even powers of $\mu$ appear in $T_{p}$ because of the U(1) charge conservation while 
in $\tilde T_{p}$ the powers are either all even or all odd depending on the value of $p$.} 
\be
T_{p}(\hat g,C; q)= \sum_{k=0}^{\infty} \mu^{2k} T_{p}^{(2k)}(C;q)\, , \qquad 
\tilde T_{p}(\hat g,C; q)= \sum_{k=0}^{\infty} \mu^{k} \tilde T_{p}^{(k)}(C;q) \, .
\ee
The su(2) Lie algebra generators are given by the U(1) charge $J_{0}^{3}$ and 
\be\label{gener}
J_{0}^{\epsilon} = \oint\! \frac{d\xi}{2\pi i }  :\! e^{i\sqrt{2}\epsilon \phi_{L}(\xi)}\!\!:  
\ee
where $\epsilon = \pm$. 
Expanding the exponential in (\ref{ghat}) and using (\ref{gener}) we represent the amplitude $T_{p}^{(2k)}(C;q)$   
in terms of nested contour integrals of correlators on an annulus $1>|z|>q$:
\bea
\label{T_amp}
 &&T^{(2k)}_{p}(C;q) = \frac{(\pi i)^k}{k!} \sum_{\{\epsilon_i\}} \oint\! \frac{d\xi_{2k}}{2\pi i } \dots \oint\! \frac{d\xi_1}{2\pi i } \nonumber \\
&&  \langle p| \prod_{n=1}^{\infty} e^{Ca_{n}\bar a_n } :\! e^{i\sqrt{2}\epsilon_{2k} \phi_{L}(\xi_{2k})}\!\!: 
\dots :\! e^{i\sqrt{2}\epsilon_1 \phi_{L}(\xi_1)}\!\!:  q^{L_{0} + \bar L_{0}}  \prod_{m=1}^{\infty} e^{a^{\dagger}_{m}\bar a_{m}^{\dagger}} |p\rangle
\eea
where the contours of integration are circles with radii
\be
1> |\xi_{2k}| > \dots > |\xi_{1}| > q \, 
\ee 
and the sum over $\{\epsilon_i\}$  goes over all distinct assignments $\epsilon_i =\pm 1$.

Normal ordering  the product of exponentials and using (\ref{int1}), (\ref{int2}) we obtain 
\bea
&& T^{(2k)}_{p}(C;q) = \frac{(\pi i)^k}{k!} \sum_{\{\epsilon_i\}} \oint\! \frac{d\xi_{2k}}{2\pi i }  \dots \oint\! \frac{d\xi_1}{2\pi i }
\prod_{i<j}^{2k}(\xi_i - \xi_j )^{2\epsilon_i \epsilon_j}   \prod_{l=1}^{2k} \xi_{l}^{\sqrt{2}p\epsilon_i}   \prod_{n=1}^{\infty} \int \frac{d^2 z}{\pi}\int \frac{d^2 w}{\pi}\nonumber \\
&& e^{-z\bar z} 
 e^{-w\bar w} \langle p| e^{ -a_n zC -\bar a_n \bar z}
\exp\Bigl[  \sum_{i=1}^{2k} \sqrt{\frac{2}{n}} a_{n}^{\dagger}\xi_i^{n}\epsilon_{i}   \Bigr]   
\exp\Bigl[  -\sum_{j=1}^{2k} \sqrt{\frac{2}{n}} a_{n} \xi_j^{-n}\epsilon_{j}   \Bigr]  e^{-a_n^{\dagger}  w q^n -\bar a_n^{\dagger} \bar w q^n } |p\rangle  
\nonumber \\
&& = \frac{(\pi i)^k}{k!} T_{k}(C;q)\sum_{\{\epsilon_i\}} \oint\! \frac{d\xi_{2k}}{2\pi i } \dots \oint\! \frac{d\xi_1}{2\pi i }
\prod_{i<j}^{2k}(\xi_i - \xi_j )^{2\epsilon_i \epsilon_j}  \prod_{l=1}^{2k} \xi_{l}^{\sqrt{2}p\epsilon_i}   \nonumber \\
&& \times \prod_{n=1}^{\infty} 
\exp\Bigl( -\frac{2Cq^{2n}\epsilon_i \epsilon_j}{n(1-Cq^{2n})} \left[  \left( \frac{\xi_{i}}{\xi_j}\right)^{n}  
+ \left( \frac{\xi_{j}}{\xi_i}\right)^{n}  \right]  \Bigr)
\eea
where 
\be\label{T2k}
T_{2k}(C;q)\equiv \prod_{n=1}^{\infty} \left( \frac{1}{1-Cq^{2n}} \right)\prod_{m=1}^{\infty} \left(  1 - q^{2m}\right) ^{4k C^m } \, .
\ee
We can also rewrite 
\be\label{inf_prod_rep}
 \prod_{n=1}^{\infty} 
\exp\Bigl[ -\frac{2Cq^{2n} \epsilon_i \epsilon_j }{n(1-Cq^{2n})}\left( \frac{\xi_{i}}{\xi_j}\right)^{n}     \Bigr]
= \prod_{m=1}^{\infty}\left(  1 - \frac{\xi_i q^{2m}}{\xi_j}\right) ^{2\epsilon_i \epsilon_j C^m } \, .
\ee

\subsection{The amplitude $T^{(2)}_{p}(C; q)$ and correction to fusion Casimir energy }\label{t22sec}

Specializing now to the first non-trivial case $k=1$ we have 
\be\label{T2C}
T^{(2)}_{p}(C;q) = -\frac{\pi^2}{2} T_{2}(C;q) I^{(2)}_{p}(C;q)
\ee
where $T_{2}(C;q)$ is defined in (\ref{T2k}) and   the contour integrals contribution is 
\bea \label{2cont}
&& I^{(2)}_{p} (C;q)= \oint\! \frac{d\xi_2}{2\pi i } \oint\! \frac{d\xi_1}{2\pi i } \frac{1}{(\xi_1-\xi_2)^2} 
\Bigl(\left(\frac{\xi_1}{\xi_2}\right)^{\sqrt{2}p} + \left(\frac{\xi_2}{\xi_1}\right)^{\sqrt{2}p} \Bigr)
\nonumber \\
&& \exp\Bigl[ \sum_{n=1}^{\infty} \frac{2Cq^{2n}}{n(1-Cq^{2n})}  \Bigl(   \left(\frac{\xi_1}{\xi_2}\right)^n + \left(\frac{\xi_2}{\xi_1}\right)^n \Bigr) \Bigr]
\eea
where we used U(1) - charge conservation to set $\epsilon_1\epsilon_2=-1$.
Expression (\ref{2cont})   simplifies to a single contour integral 
\be\label{1cont}
I^{(2)}_{p}(C;q)=\frac{1}{2\pi i } \oint\!\frac{d\xi}{(1-\xi)^2}\exp\Bigl( \sum_{n=1}^{\infty} \frac{2Cq^{2n}}{n(1-Cq^{2n})}  
[  \xi^n + \xi^{-n} ] \Bigr)\left( \xi^{\sqrt{2}p} +  \xi^{-\sqrt{2}p} \right)
\ee
where the contour is a circle centred at the origin with  a radius $|\xi|$ such that  $q^2<|\xi|<1$.

The leading singularity of $T^{(2)}_{p}(C;q)$ at $q\to 1$ is of the form 
\be
T^{(2)}_{p}(C;q)\sim \frac{A(C)}{1-q^2} e^{-\frac{{\cal E}_{D}}{\epsilon}} \sim \frac{A(C)}{2\epsilon} e^{-\frac{{\cal E}_{D}}{\epsilon}}
\ee
where the residue $A(C)$ gives the correction to the fusion Casimir energy $\Delta {\cal E}_{0} =\frac{1}{2}\mu^2A(C)$ (\ref{DCas2}).
As shown in appendix \ref{appA} this correction is 
\be
\Delta {\cal E}_{0}  = \frac{1}{2}\mu^2  \pi^2f(C) {\cal F}(C)\sqrt{1-C}     
\ee
where 
\be
f(C) = \frac{\Gamma(2\delta+1)}{2^{1+2\delta}\Gamma(\delta)\Gamma(\delta+2)}\, 
{}_{2}F_{1}\left(2,2\delta+1; \delta+2;\frac{1}{2}\right) \, ,
\ee
\be
\delta = \frac{2C}{1-C} \, , 
\ee
and 
\be
{\cal F}(C) = \exp\Bigl[ \frac{4C}{1-C}\sum_{m=1}^{\infty}C^{m}\ln\left(1+\frac{1}{m}\right)\Bigr] \, .
\ee
This correction is second order in $\mu$ and contains all orders in $C$. At the order $\mu^2 C$ it matches with 
(\ref{fusion_pert}).  We note that $f(C) {\cal F}(C)/\delta >0$ so that $\Delta {\cal E}_{0}$ 
has the same sign as $C$. This sign is opposite to that of ${\cal E}_{D}$. Hence
at the leading order the Casimir energy is shifted towards ${\cal E}_{N}$. At higher orders we expect to get some  function interpolating between 
${\cal E}_{D}$ 
and ${\cal E}_{N}$.   Unfortunately we do not know how  to calculate this function non-perturbatively. 
                             Note that the leading correction ${\cal E}_{0} $ came out to be independent of the momentum $p$. This means that 
                             (at least for small $\mu$) a 
                          single   multiplicative counterterm (as in \ref{fuse_naive}) will retain all Dirichlet-type Ishibashi states. 
                             

\subsection{Resummed RG logarithms}
The RG logarithms that appear in the $C$-expansion (\ref{Cexp}) are also contained in the terms $T^{(2k)}_{p}(C;q)$. 
In particular we checked\footnote{The easiest way to do it is by using the infinite product representation (\ref{inf_prod_rep}). 
} that we obtain the leading logarithm present in (\ref{llog}) at the order $\mu^2$ by expanding $T^{(2)}_{0}(C;q)$ to the order $C^2$.

Another interesting quantity in which we can observe how the RG logarithms are resummed in power functions is 
$\tilde T_{\frac{1}{\sqrt{2}}}(\hat g, C;q)$. The leading logarithm it contains was calculated in (\ref{charge_log}): 
\be\label{charge_log2}
\tilde T_{\frac{1}{\sqrt{2}}}^{(1)} (\hat g ; q)\sim 2\pi i \mu \ln(1-q) \, .
\ee

We can also calculate $\tilde T_{\frac{1}{\sqrt{2}}}(\hat g, C;q)$ to the first order in $\mu$ and to all orders in $C$. We have 
\be
\tilde T_{\frac{1}{\sqrt{2}}}^{(1)}( C;q) = \left\langle \frac{1}{\sqrt{2}}\right| \prod_{n=1}^{\infty}e^{Ca_n\bar a_n}\oint \frac{d\xi}{2\pi i } 
:\! e^{i\sqrt{2}\phi_{L}(\xi)}\!: \prod_{m=1}^{\infty} e^{q^{2m}a_{m}^{\dagger}\bar a_{m}^{\dagger}}\left|-\frac{1}{\sqrt{2}}\right\rangle
\ee
Using (\ref{int1}), (\ref{int2}) we find 
\be
\tilde T_{\frac{1}{\sqrt{2}}}^{(1)}(C;q) =\left(\prod_{n=1}^{\infty}\frac{1}{1-Cq^{2n}}\right)\prod_{m=1}^{\infty} \exp\left( -\frac{2Cq^{2m}}{m(1-Cq^{2m})}\right)  
\, .
\ee
The first factor here contains the essential singularity at $q=1$ that corresponds to the fusion Casimir energy (\ref{DCas1}), (\ref{DCas2}). The second factor 
(assuming $C\ne 1$) has a power singularity at $q=1$: 
\be\label{factor}
\prod_{m=1}^{\infty} \exp\left( -\frac{2Cq^{2m}}{m(1-Cq^{2m})}\right)  \sim (1-q^2)^{\frac{2C}{1-C}}
\ee
Expanding this power function in series in $C$ we recover at the leading order the logarithm  (\ref{charge_log2}). Hence the amplitudes $T_{p}(\hat g, C;q)$ 
contain  RG logarithms resummed into power functions.  

Same methods that we used to analyse $T^{(2)}_{0}(C;q)$ (see section \ref{t22sec} and Appendix A) 
can be used to  calculate the asymptotic for  $\tilde T_{\sqrt{2}}^{(2)}$ which we give here 
without details  
\be
\tilde T_{\sqrt{2}}^{(2)} \sim (1-q^2)^{\frac{4C}{1-C}} e^{-\frac{{\cal E}_{D}}{\epsilon}} \qquad q\to 1 \, .
\ee

We see that unlike the leading corrections $T_{p}^{(2)}(C;q)$ for the Dirichlet type amplitudes which contain a simple pole, for 
the Neumann type amplitudes we get irrational power functions. Although we could not calculate asymptotics for any of the next corrections 
$\tilde T^{(3)}(C;q)$ we tried to glean more information from the next-to-leading order $C$-expansion amplitude  $\tilde T^{(2)}_{\frac{1}{\sqrt{2}}}(\hat g;q)$ 
presented in Appendix B. 
We found that a power function different from the leading order must be present in  $\tilde T^{(3)}(C;q)$. This leaves us with no good guess for 
the general form of the predexponential function $A_{p}(\epsilon)$ in the asymptotics 
\be
\tilde T_{p}(\hat g, C;q) \sim A_{p}(\epsilon)e^{-\frac{{\cal E}^{\rm N}_{p}}{\epsilon}}\, ,  \qquad q\to 1 \, . 
\ee

  Of course by T-duality the behaviour of $\tilde T_{p}$ amplitudes near the Neumann brane $\mu=1/2$ is the same (up to switching the sign of $C$) 
  as that of the Dirichlet type amplitudes $T_{p}$ near $\mu=0$.

\section{$C=1$}\label{C1}
\setcounter{equation}{0}
Calculating the fusion boils down to calculating the amplitudes 
$$
T_{p}(\hat g;q,C)= {\rm Tr}(  \Pi_{p}q^{2L_0} C^{N} \hat g)
$$ discussed  after equation 
 (\ref{basic_amp}). Although for  generic value of  $C$ we are not aware of a method to calculate this quantity, we can do so at $C=1$ where the additional 
 weighting by the oscillator number $N$ disappears.
 The value $C=1$ formally corresponds to decompactifying the free boson, i.e. taking the radius 
  $R=\infty$. The states with $M\ne 0 $ have  divergent conformal weights while the states with $M=0$ and a finite $N$ all tend to the vacuum state. 
  To obtain a state with nonzero momentum $p$ in ${\cal H}_{\infty}$ one would need to take states $|N,0\rangle_{R}$ with increasing $N$ in such a way that 
  the ratio $N/{R}$ tends to $p$ in the limit $R_1\to \infty$. Thus, the only state that remains intact when passing to infinite radius is the vacuum. 
The related vacuum amplitude 
 $T_{0}(\hat g;q,1)$ can be written as 
\be
Z(t,\mu)\equiv T_{0}(\hat g;q,1) = \frac{1}{2\pi} \int\limits_{0}^{2\pi}\!\! dx 
\, {\rm Tr}\left( \hat g\, e^{2ix J_{0}^{3}} q^{2L_{0}}\right) 
\ee
where the trace is taken over  ${\cal H}^{\rm chiral}$ given in (\ref{Hchiral}) and $q=e^{-2\pi t}$. 
As before we take $\hat g$ to be given by (\ref{ghat}), (\ref{g}). 

The operator $\hat g e^{2ix J_{0}^{3}}$ corresponds to a group element 
\be
\left( \begin{array}{rr}
\cos(\pi\mu)e^{ix}& i \sin(\pi \mu) e^{-ix}\\
i\sin(\pi\mu)e^{ix}& \cos(\pi\mu)e^{-ix}
\end{array}\right) \, .
\ee
This element can be diagonalised by the adjoint action of a  suitable group element $h$ to give 
\be
h (ge^{2ix J_{0}^{3}}) h^{-1} = \left(\begin{array}{cc}
e^{i\varphi}& 0\\
0&e^{-i\varphi}
\end{array}\right) 
\ee 
  where 
  \be
  \cos(\varphi) = \cos(x)\cos(\pi\mu) \, .
  \ee
   Computing the trace in this basis we obtain 
   \be\label{Z2}
   Z(t,\mu)=\frac{1}{2\pi \eta(2it)}\int\limits_{0}^{2\pi}\!\! dx \sum_{n=-\infty}^{\infty}e^{in\varphi}q^{n^2/2}
   \ee
  where 
  \be
  \eta(2it)  = q^{\frac{1}{12}}\prod_{n=1}^{\infty} (1-q^{2n}) 
  \ee
  is the Dedekind eta function. We further rewrite  (\ref{Z2}) in terms of Chebyshev polynomials of the first 
  kind
  \be
  Z(t,\mu) = \frac{1}{\pi \eta(2it)}\int\limits_{0}^{2\pi}\!\! dx\,  \Bigl[\sum_{n=0}^{\infty} q^{n^2/2}T_{n}(\cos(\pi\mu)\cos(x)) -\frac{1}{2}\Bigr]\, .
  \ee
  Using the generating function 
  \be
  \sum_{n=0}^{\infty}T_{n}(\xi)p^n = \frac{1-p\xi}{1-2p\xi + p^2} 
  \ee
  setting $p=e^{iy}$ and using 
  \be
  \int\limits_{-\infty}^{\infty} e^{iny}e^{-\frac{y^2}{4\pi t}} dy = 2\pi \sqrt{t} e^{-\pi t n^2} = 2\pi \sqrt{t} q^{n^2/2}
  \ee
  we further rewrite $Z(t,\mu)$ as 
  \be\label{ZZ}
  Z(t,\mu) = \frac{1}{4\pi^2\sqrt{t}\eta(2it)}\int_{-\infty + i\epsilon}^{\infty + i\epsilon} \!\!dy\, 
  e^{-\frac{y^2}{4\pi t} }\int\limits_{0}^{2\pi}\!\!dx\, \frac{1-e^{2iy}}{1-2e^{iy}\cos(\pi\mu)\cos(x) + e^{2iy}} \, .
  \ee
  The $x$-integral can be now taken via residues. After some tedious but straightforward calculations 
  we obtain\footnote{An alternative way to obtain (\ref{Z3}) is to use modular transformation for the theta function in 
  (\ref{Z2}) and then use periodicity to change the integration variable to $\varphi$. The $i\epsilon$ regularisation would then have to be 
  introduced by hand.} 
  \be\label{Z3}
  Z(t,\mu) = \frac{1}{2\pi \sqrt{t}\eta(2it)}\sum_{n=-\infty}^{\infty} 
  \int\limits_{\pi(n+\mu)+i\epsilon}^{\pi(n+1-\mu)+i\epsilon} \!\!dy\, e^{-\frac{y^2}{4\pi t} } \frac{|\sin(y)|}{\sqrt{\cos^2(\pi \mu) 
  - \cos^2(y)}} \, .
  \ee
  At the Dirichlet point $\mu=0$ we have 
  \be
  Z(t,0) = \frac{1}{\eta(2it)}
  \ee
  while at the Neumann point $\mu=1/2$ the $i\epsilon$ prescription in (\ref{ZZ}) gives us 
  \bea\label{ZN}
  Z(t,1/2) && = \frac{1}{2\pi \sqrt{t}\eta(2it)}(-i)\int\limits_{-\infty + i\epsilon}^{\infty + i\epsilon}\!\!dy\, e^{-\frac{y^2}{4\pi t} } \tan(y) 
  \nonumber \\ && = \frac{1}{\sqrt{t}\eta(2it)}\sum_{n=0}^{\infty} \exp\left( -\frac{\pi(n+1/2)^2}{4t}\right) 
   = \frac{\vartheta_{10}(0,\tau)}{2\sqrt{t}\phi(q^2)} \, .
  \eea
  Using the identity 
  \be
  \prod_{m=1}^{\infty} (1+p^m)(1-p^{2m-1}) = 1 
  \ee
  we recast (\ref{ZN}) as 
  \be
  Z(t,1/2) = q^{-\frac{1}{12}}\prod_{n=1}^{\infty}\frac{1}{1+q^{2n}} 
  \ee
  which is the expected result for the Neumann brane. 
  
  For  intermediate values $0<\mu<1/2$ we can obtain the $t\to 0$ asymptotic of (\ref{Z3}) using the saddle point approximation. 
  The leading contribution to the integral in (\ref{Z3}) comes from small regions near $y=-\pi\mu$ and $y=\pi\mu$. 
  We have 
  \be
  Z(t,\mu) \sim \frac{1}{\pi\sqrt{t}\eta(2it)}\int\limits_{\pi\mu}^{\infty}\!\!\frac{dy}{\sqrt{y-\pi\mu}}e^{-\frac{y^2}{4\pi t} } 
  = \frac{1}{\pi\sqrt{t}\eta(2it)}\sqrt{\frac{\pi \mu}{2}} e^{-\frac{\pi \mu^2}{8 t} } K_{\frac{1}{4}}\left(\frac{\pi \mu^2}{8 t} \right)
  \ee
  where $K_{1/4}$ stands for a modified Bessel function. Using the asymptotics 
  \be
  K_{\alpha}(z) \sim \sqrt{\frac{\pi}{2z}}e^{-z} \, , \qquad z\to \infty 
  \ee
  and 
  \be
  \frac{1}{\eta(2it)} \sim \sqrt{2t}\displaystyle{e^{\frac{\pi}{24 t}}} \, , \qquad t \to 0 
  \ee
  we finally obtain 
  \be\label{Z_asympt}
  Z(t,\mu) \sim  \sqrt{\frac{2t\sin(\pi\mu)}{\pi\mu\cos(\pi\mu)}} \exp\left( \frac{\pi}{4t}(\frac{1}{6}-\mu^2 ) \right) 
  \, , \quad t\to 0 \, . 
  \ee
  While the exponent interpolates continuously between the Dirichlet and Neumann Casimir energies ${\cal E}_{D}$ and ${\cal E}_{N}$ 
  as we vary $\mu$, the predexponential factor blows up at $\mu=1/2$ that reflects a discontinuity: the limits $\mu\to 1/2$ 
  and $t\to 0$ do not commute. Apart from the $\mu$-dependent constant prefactor\footnote{ We have verified the ${\cal O}(\mu^2)$ 
  term in (\ref{ff}) by an independent perturbative calculation the details of which we omit. } 
  \be\label{ff}
  f(\mu) = \sqrt{\frac{\sin(\pi\mu)}{\pi\mu\cos(\pi\mu)}} = 1 + \frac{\pi^2 \mu^2}{6} + \frac{19\pi^4\mu^4}{360} 
  + {\cal O}(\mu^6) 
    \ee
    and the $\mu$-dependent Casimir energy, 
  the $t\to 0$ singularity is qualitatively the same for all values $0\le \mu <1/2$. For all of these values the $\sqrt{t}$ 
  vanishing factor is present. The value $\mu=1/2$ is special in that the $\sqrt{t}$ factor is absent. 
    
 This discontinuity is qualitatively the same as the one we expect from the induced boundary RG flow discussed in the introduction: 
 all branes with $0\le \mu <1/2$  flow to the Dirichlet brane while the Neumann brane ($\mu = 1/2$) is transported to the Neumann brane 
 at the new radius. 
  The presence of the $f(\mu)$   factor is consistent with the conclusion made in section \ref{vacuum_sing2} that the fusion gives 
  a Dirichlet or Neumann boundary state multiplied by a non-trivial function of $\mu$ and $C$.

 \section{Concluding remarks} \label{conclusions}  
   In this paper we studied the fusion of a radius changing interface with an exceptional D-brane. Our motivation was twofold - 
   to make connection with bulk induced boundary RG flows and to get an insight into the fusion singularity structure in a symmetry 
   breaking situation when topological defect methods seem to be of no use. 
   
   The radius changing interface depends on the radius related parameter $C$ while the exceptional boundary state is parameterised 
   by another parameter --  $\mu$.  
   We have developed two different perturbative expansions in which one of the two parameters is treated non-perturbatively. 
   The fusion process is essentially non-perturbative so that perturbative calculations are of limiting value. We did observe however 
   how RG logarithms occur in the fusion process and how they get resummed into power singularities. This confirms our conjecture 
   that the fusion describes the corresponding bulk induced boundary RG flow in a particular renormalisation scheme. 
   We discussed potential ambiguities present in the induced flows in the introduction around formula (\ref{beta_general2}). 
   We also calculated the vacuum fusion amplitude non-perturbatively for $C=1$ that corresponds to infinite radius deformation. This  
   amplitude exhibits  the same type of discontinuity as observed for the induced RG flows of \cite{Gaberdiel_etal}.

   A complete non-perturbative control over fusion would be possible if one could calculate the traces $T_{p}(\hat g;q,C)$ 
   defined in    (\ref{basic_amp}) as traces of certain operators in a chiral Fock space. We hope to make progress on this in the future. 
   Among these amplitudes the vacuum fusion amplitude $T_{0}(\hat g;q,C)$ is special as we do know what singularity to expect in it. 
   It also contains information on the $g$-factor. 
   At first glance, gaining control over this  amplitude  could potentially be a stronger tool than $g$-theorem, 
   as  we would hope to get a prediction for the actual value of the infrared $g$-factor. However the infrared $g$-factor  appears to 
   be masked by an extra finite multiplicative renormalisation which emerges in the fusion process as  discussed in section \ref{vacuum_sing2}
   and after formula (\ref{ff}).


\begin{center}
{\bf Acknowledgments}
\end{center}

The author is grateful to Daniel Friedan and Robert Weston for stimulating discussions. 
This work    was supported in part by STFC grant "Particle Theory at the Higgs Centre", ST/L000334/1.


\appendix
\renewcommand{\theequation}{\Alph{section}.\arabic{equation}}
\setcounter{equation}{0}

\section{ Correction to fusion Casimir energy  }\label{appA}

For $-1<C<1$ the most singular part of (\ref{1cont}) at $q\to 1$ can be obtained by replacing $I_{p}^{(2)}(C;q)$ by $\tilde I_{p}^{(2)} + \tilde I_{-p}^{(2)}$
where 
\be
\tilde I^{(2)}_{p}(C;q)=\frac{1}{2\pi i }  \oint\!\frac{d\xi}{(1-\xi)^2}\left( 1 - q^2 \xi\right)^{-\delta}\left( 1 - \frac{q^2}{\xi}\right)^{-\delta}\xi^{\sqrt{2}p} \, , 
\ee
\be
\delta = \frac{2C}{1-C} \, 
\ee
and there is a branch cut on the real line extending from $\xi=0$ to $\xi = q^2$. 
Expanding 
\be
\frac{1}{(1-\xi)^2} = q^4 \sum_{n=0}^{\infty} (n+1)(1-q^2)^n(1-q^2\xi)^{-2-n}
\ee
and taking the contour integral we obtain the expansion  
\bea \label{II1}
&& \tilde I^{(2)}_{p}(C;q) = \frac{\Gamma(N+1+\delta)}{\Gamma(\delta)(N+1)!} q^{6+2N} \nonumber \\
&& \times\left(\sum_{n=0}^{\infty}(n+1)(1-q^2)^{n}\, {}_2F_1(1+N+ \delta, 2+n+\delta;2+N ;q^4)\right)  \,  
\eea
when $p=\frac{N}{\sqrt{2}}>0 $, 
and 
\be\label{II2}
 \tilde I^{(2)}_{p}(C;q) = q^{2+2N} 
 \sum_{n=0}^{\infty}(n+1)(1-q^2)^{n}\frac{\Gamma(N+\delta + n + 1)}{\Gamma(2+\delta + n)}\, 
{}_2F_1( \delta, N+\delta+n+1;N ;q^4) \,  
\ee
when $p=\frac{N}{\sqrt{2}}<0 $. 

Using the transformation 
\bea
 && {}_{2}F_1(a,b;c;z) = \frac{\Gamma(c)\Gamma(c-a-b)}{\Gamma(c-a)\Gamma(c-b)}\, {}_{2}F_{1}(a,b;a+b+1-c;1-z) \nonumber \\
&& + \frac{\Gamma(c)\Gamma(a+b-c)}{\Gamma(a)\Gamma(b)}(1-z)^{c-a-b}\, {}_{2}F_{1}(c-a,c-b;1+c-a-b;1-z) 
\eea
for each term in the above expansions (\ref{II1}), (\ref{II2}) we find the asymptotic 
\be\label{I2sing}
I^{(2)}_{p}(C;q) \sim 2f(C) (1-q^2)^{-1-2\delta} \quad \mbox{ for } q\to 1
\ee
where 
\bea
f(C) &=& \frac{1}{\Gamma(\delta)} \sum_{n=0}^{\infty}\frac{(n+1)\Gamma(n+1+2\delta)}{\Gamma(2+n + \delta)}\left(\frac{1}{2}\right)^{n+1+2\delta}
\nonumber \\ 
& =&
\frac{\Gamma(2\delta+1)}{2^{1+2\delta}\Gamma(\delta)\Gamma(\delta+2)}\, 
{}_{2}F_{1}\left(2,2\delta+1; \delta+2;\frac{1}{2}\right) \, . 
\eea
We further calculate the asymptotic 
\be 
\prod_{m=1}^{\infty} \left(  1 - q^{2m}\right) ^{4 C^m } = \exp\left[ -4C\sum_{n=1}^{\infty}\frac{q^{2n}}{n(1-Cq^{2n})} \right] 
\sim (1-q^2)^{2\delta} {\cal F}(C) 
\ee
where 
\be\label{FF}
{\cal F}(C) = \exp\Bigl[ \frac{4C}{1-C}\sum_{m=1}^{\infty}C^{m}\ln\left(1+\frac{1}{m}\right)\Bigr] \, .
\ee

Putting together (\ref{T2C}), (\ref{I2sing}), (\ref{FF})  we obtain the leading singularity 
\be
T^{(2)}_{p}(C;q) \sim  - \frac{\pi^2 f(C) {\cal F(C)} }{1-q^2} \prod_{n=1}^{\infty} \left( \frac{1}{1-Cq^{2n}} \right)\, . 
\ee
Comparing this with (\ref{DCas1}), (\ref{DCas2})   we find the following correction to the fusion Casimir energy 
\be
\Delta {\cal E}_{0} = \mu^2  \pi^2f(C) {\cal F(C)}\sqrt{1-C}
\ee

\section{The amplitude  $\tilde T^{(2)}_{\frac{1}{\sqrt{2}} }(\hat g;q)$}
\setcounter{equation}{0}
We have 
\be
 \tilde T^{(2)}_{\frac{1}{\sqrt{2}}}(\hat g;q) = \sum_{n<m} \left\langle \frac{1}{\sqrt{2}}\right| q^{2(n+m)} a_n a_m \hat g a_{n}^{\dagger}a_{m}^{\dagger} 
\left|-\frac{1}{\sqrt{2}}\right\rangle 
+ \frac{1}{2!}\sum_{n=1}^{\infty}\left\langle \frac{1}{\sqrt{2}}\right| q^{4n} (a_n)^2 \hat g (a_n^{\dagger})^{2}\left|-\frac{1}{\sqrt{2}}\right\rangle \, . 
\ee
A straightforward calculation gives 
\bea
&& \sum_{n<m} \left\langle \frac{1}{\sqrt{2}  }\right| q^{2(n+m)} a_n a_m \hat g a_{n}^{\dagger}a_{m}^{\dagger} 
\left|-\frac{1}{\sqrt{2}}\right\rangle = a_1 q^4\left( \frac{1}{(1-q^2)(1-q^4)} - 1\right) \nonumber \\ 
&& +  (a_3 - 2a_2) q^2 \frac{\ln(1-q^4)}{1-q^2}  -a_3 \left(q^2\frac{\ln(1+q^2)}{1-q^2}  - q^4 \right) \nonumber \\
&& + (a_3-\frac{a_2}{2})  ( [\ln(1-q^2)]^2 - {\rm Li}_{2}(q^4)) \,  , 
\eea
\bea
&& \sum_{n=1}^{\infty}\left\langle \frac{1}{\sqrt{2}}\right| q^{4n} (a_n)^2 \hat g (a_n^{\dagger})^{2}\left|-\frac{1}{\sqrt{2}}\right\rangle = 
a_1 \frac{q^4}{1-q^4}  + (2 a_3 - a_2)  \ln(1-q^4) \nonumber \\
&& + (a_3 - a_2) {\rm Li}_{2}(q^4) \, 
\eea
where 
\be
a_1 = i\sin(\pi\mu) \cos^2(2\pi\mu)\, , 
\ee
\be
a_2 = i\sin(\pi \mu)\sin^2(2\pi \mu)\, , 
\ee
\be
a_3 = i\sin(2\pi \mu)\cos(2\pi \mu)\cos(\pi \mu)  \, .
\ee


\end{document}